\documentclass[journal,10pt,onecolumn,doublespace,draftclsnofoot,]{IEEEtran}
\usepackage{amssymb}
\usepackage[figuresright]{rotating}
\usepackage{amsmath,graphicx,epsfig,latexsym,mathrsfs,epsf,bm}
\usepackage{amsthm}
\usepackage{color}
\usepackage{epstopdf}
\def\Pr{\mathop{\textrm{Pr}}}

\newsavebox{\fminibox}
\newlength{\fminilength}

  \def\+{^\dagger}

\def\nequiv{\not\kern-.05em\equiv}
\def\egal{\kern-.5em=\kern-.5em}        
\def\propt{\kern-.2em\propto\kern-.2em} 

\def\argmin{\mathop{\mathrm{arg\,min}}} 
\def\intdouble{\int\kern-0.3em\int}
\def\inttriple{\int\kern-0.3em\int\kern-0.3em\int}

\def\rond#1{\overset{\kern-0.33em~_\circ}{#1}}
\def\rondit[#1]#2{\overset{\kern#1~_\circ}{#2}}

\usepackage{subcaption}
\usepackage{mathabx}
\usepackage{xfrac}
\usepackage{upgreek}

\usepackage{tikz}
\usetikzlibrary{shapes,arrows,snakes,positioning}
\usepackage{circuitikz}
\usetikzlibrary{arrows,automata}
\usepackage[latin1]{inputenc}

\newtheorem{Theorem}{Theorem}

\newtheorem{Corollary}[Theorem]{Corollary}
\newtheorem{CounterExample*}{$\overline{\hbox{\bf Example}}$}
\newtheorem{Definition}{Definition}

\newtheorem{Example*}{Example}

\newtheorem{Intuition*}{Intuition}
\newtheorem{Joke*}{Joke}
\newtheorem{Lemma}[Theorem]{Lemma}
\newtheorem{Lemma*}{Lemma}

\newtheorem{Note*}{Note}
\newtheorem{Open problem}{Open problem}
\newtheorem{Proposition}{Proposition}

\newtheorem{Question*}{Question}
\newtheorem{Remark*}{Remark}

\newcommand{\uXM}{\underline{{X}}_{\mathcal{M}}}

\newcommand{\tPr}{\textnormal{Pr}}

\newcommand{\exc}{{\rm  E}}
\newcommand{\mcN}{{\mathcal N}}
\newcommand{\mcM}{{\mathcal M}}
\newcommand{\mcE}{{\mathcal E}}

\newcommand{\mcT}{{\mathcal T}}
\newcommand{\mcH}{{\mathcal H}}

\newcommand{\mcV}{{\mathcal V}}
\newcommand{\mcD}{{\mathcal D}}
\newcommand{\mcG}{{\mathcal G}}
\newcommand{\mcX}{{\mathcal X}}

\newcommand{\mcJ}{{\mathcal J}}

\newcommand{\mbbR}{{\mathbb R}}

\newcommand{\uzm}{\underline{0}}
\newcommand{\ux}{\underline{{x}}}
\newcommand{\ue}{\underline{e}}
\newcommand{\uW}{\underline{W}}
\newcommand{\uZ}{\underline{Z}}

\newcommand{\uV}{\underline{V}}
\newcommand{\uX}{\underline{{X}}}

\newcommand{\bSigma}{\mathbf{\Sigma}}

\newcommand{\bDelta}{\mathbf{\Delta}}

\newcommand{\bI}{{\bf I}}

\newcommand{\bK}{{\bf K}}
\newcommand{\bLambda}{{\bf \Lambda}}

\author{\IEEEauthorblockN{Navid Tafaghodi Khajavi and Anthony Kuh\footnote{This paper was presented for the special case of tree approximation in part at 2016 Information Theory and Application Workshop \cite{ITA2016}.}\\}
\IEEEauthorblockA{Department of Electrical  Engineering\\
University of Hawaii, Honolulu, HI 96822\\
Email: \{navidt, kuh\}@hawaii$.$edu}}

\begin{document}
\title{The Quality of the Covariance Selection Through Detection Problem and AUC Bounds}
\maketitle

\begin{abstract}
We consider the problem of quantifying the quality of a model selection problem for a graphical model.
We discuss this by formulating the problem as a detection problem. 
Model selection problems usually minimize a distance between the original distribution and the model distribution.
For the special case of Gaussian distributions, the model selection problem simplifies to the covariance selection problem which is widely discussed in literature by Dempster \cite{dempster}. To compute the model covariance matrix in \cite{dempster}, the likelihood criterion is maximized or equivalently the Kullback-Leibler (KL) divergence is minimized.
While this solution is optimal for Gaussian distributions in the sense of the KL divergence, it is not optimal when compared with other information divergences and criteria such as Area Under the Curve (AUC). 

In this paper, we present a simplified integral to numerically evaluate AUC. We analytically compute upper and lower bounds for the AUC and discuss the quality of model selection problem using the AUC and its bounds as an accuracy measure in detection problem.
We define the correlation approximation matrix (CAM) and show that analytical computation of the KL divergence, the AUC and its bounds only depend on the eigenvalues of CAM. We also show the relationship between the AUC, the KL divergence and the ROC curve by optimizing with respect to the ROC curve.
In the examples provided, we pick tree structures as the simplest graphical models. We perform simulations on fully-connected graphs and compute the tree structured models by applying the widely used Chow-Liu algorithm \cite{chowliu}. 
Examples show that the quality of tree approximation models are not good in general based on information divergences, the AUC and its bounds when the number of nodes in the graphical model is large.
Moreover, we show both analytically and by simulations that the $1-$AUC for the tree approximation model decays exponentially as the dimension of graphical model increases.

\end{abstract}

\section{Introduction}
\label{sec:intro}

Graphical models are useful tools for describing the geometric structure of networks in numerous applications such as energy, social, sensor, neuronal, and transportation networks \cite{SPOGraph} that deal with high dimensional data.
Learning from these high dimensional data requires large computation power which is not always available \cite{kollerPGM}, \cite{jordanLGM}.
The hardware limitation for different applications force us to compromise between the accuracy of the learning algorithm and its time complexity by using the best possible approximation algorithm and imposing structures, instead.
In other words, the main concern is to compromise between model complexity and its accuracy by choosing a simpler, yet informative model.
There are lots of approximation algorithms that are proposed for model selection to impose structure given data.
For the Gaussian distribution, the covariance selection problem is presented and studied in \cite{dempster} and \cite{GMbook96}.

The ultimate purpose of the covariance selection problem is to reduce the computation complexity in various applications.
One of the special approximation models is the tree approximation model. Tree approximation algorithms are among the algorithms that reduce the number of computations to get quicker approximate solutions to a variety of problems.
If a tree model is used, then distributed estimation algorithms such as message passing algorithm \cite{BP2} and the belief propagation algorithm \cite{BP} can easily be applied and are guaranteed to converge to the maximum likelihood solution.

There are algorithms in the literature such as the Chow-Liu minimum spanning tree (MST) \cite{chowliu}, the first order Markov chain approximation \cite{ISIT2013} and penalized likelihood methods such as LASSO \cite{LASSO} and graphical LASSO \cite{GLASSO} that can be used to approximate the correlation matrix and the inverse correlation matrix with a more sparse graph representation while retaining good accuracy. 
The Chow-Liu MST algorithm for Gaussian distribution is to find the optimal tree structure using a Kullback-Leibler (KL) divergence cost function \cite{dempster}. The Chow-Liu MST algorithm constructs a weighted graph by computing pairwise mutual informations and then utilizes one of the MST algorithms such as the Kruskal algorithm \cite{kruskal} or the Prim algorithm \cite{prim}.
The first order Markov chain approximation uses a regret cost function to output first order Markov chain structured graph \cite{ISIT2013} by utilizing a greedy type algorithm. 
Penalized likelihood methods uses an L1-norm penalty term in order to sparsify the graph representation and eliminating some of the edges.
Recently, tree approximation in a linear, underdetermined model is proposed in \cite{ICASSP17} where the solution is based on expectation, maximization (EM) algorithm combined with the Chow Liu algorithm.

Sparse modeling has many applications in distributed signal processing and machine learning over graphs. One of its applications is in {\it smart grid}.
Smart grid is a promising solution that delivers reliable energy to consumers through the power grid when there are uncertainties such as distributed renewable energy generation sources.
Smart grid technologies such as smart meters and communication links are added to the power grid in order to obtain the high dimensional, real-time data and information ``{\it Big Data}," and overcome uncertainties and unforeseen faults.
The future grid will incorporate distributed renewable energy generation such as solar PVs, with these sources
being highly correlated. 
Thus, modeling is an essential part for signal processing and implementation of the smart grid.

We discuss the quality of the model selection problem, focusing on the Gaussian case, i.e. covariance selection problem.
We ask the following important question: {\it ``is the covariance approximation of the covariance matrix for the Gaussian model a good approximation?"}
To answer this question, we need to pick a closeness criterion which has to be coherent and general enough to handle a wide variates of problems and also have asymptotic justification \cite{Method4MS}.
In many applications the Kullback-Leibler (KL) divergence has been proposed as a closeness criterion between the original distribution and its model approximation distribution \cite{dempster} and \cite{chowliu}.
Besides that, other closeness measures and divergences are used for the model selection problem. One example is the use of reverse KL divergence as the closeness criterion in variational methods to learn the desired approximation structure \cite{mackayITILA}.

In this paper we bring a different perspective to the model approximation problem by formulating a general detection problem.
The detection problem leads to calculation of the log-likelihood ratio test (LLRT) statistic, the receiver operating characteristic (ROC) curve, the KL divergence and the reverse KL divergence as well as the area under the curve (AUC) where the AUC is used as the accuracy measure for the detection problem.
The detection problem formulation gives us a broader view as well as looking at different approaches of determining whether a particular model is a good approximation or not.
More specifically, the AUC gives us additional incite about any approximation since it is a way to formalize the model approximation problem.
For Gaussian data, the LLRT statistic simplifies to an indefinite quadratic form.
A key quantity that we define is the correlation approximation matrix (CAM) as the product of the original correlation matrix and the inverse of the model approximation correlation matrix. For Gaussian data this matrix contains all the information needed to compute the information divergences, the ROC curve and the area under this curve, i.e. the AUC.
We also show the relationship between the CAM, the AUC and the Jeffreys divergence \cite{jeffreysD}, the KL divergence and the reverse KL divergence. 
We present an analytical expression to compute the AUC for a given CAM that can be efficiently evaluated numerically. We show the relation between the AUC, the KL divergence, the LLRT statistics and the ROC curve. We also present analytical upper and lower bounds for the AUC which are only depend on eigenvalues of the CAM.
Throughout the discussion section, we pick the tree approximation model as a well-known subset of all graphical models.
The tree approximations is considered since they are widely used in literature and it is much simpler performing inference and estimation on trees rather than graphs that have cycles or loops.
We perform simulations over synthetic and real data for several examples to explore and discuss our results.
Simulation results indicate that $1-$AUC is decreasing exponentially as the graph dimension increases which is consistent with analytical results obtained from the AUC upper and lower bounds.

The rest of this paper is organized as follows. In section \ref{sec:prob} we give a general framework for the detection problem and the corresponding sufficient test statistic, the log-likelihood ratio test. Moreover, the sufficient test statistic for Gaussian data as well as its distribution under both hypotheses are also presented in this section.
The ROC curve and the AUC definition as well as analytical expression for the AUC are given in section \ref{sec:AUC}. Section \ref{sec:bound} provides analytical lower and upper bounds for the AUC as function of the CAM. Moreover, Section \ref{sec:examples} presents the tree approximation model and provides some simulations over synthetic examples as well as real solar data examples and investigates quality of tree approximation based on the numerically evaluated AUC and also its analytical upper and lower bounds.
Finally, Section \ref{sec:con} summarizes results of this paper.

\section{Detection Problem Framework}
\label{sec:prob}

In this section, we present a framework to quantify the quality of a model selection problem. More specifically, we formulate a detection problem to distinguish between the covariance matrix of a multivariate normal distribution and an approximation of the aforementioned covariance matrix based on the given model.

\subsection{Model selection problem}
We want to approximate a multivariate distribution by the product of lower order component distributions \cite{lewis_approx}.
Let random vector $\uX \in \mbbR^n$, have a distribution with parameter $\Theta$, i.e. $\uX \sim f_{\uX}(\ux)$. 
We want to approximate the random vector $\uX$, with another random vector associated with the desired model\footnote{Examples of possible models: tree structure, sparse structure and Markov chain.}.
Let the model random vector $\uXM \in \mbbR^n$ have a distribution with parameter $\Theta_\mcM$, associated with the desired model, i.e. $\uX \sim f_{\uX_\mcM}(\ux)$. Also, let $\mcG=(\mcV, \mcE_\mcM)$ be the graph representation of the model random vector $\uXM$ where sets $\mcV$ and $\mcE_\mcM$ are the set of all vertices and the set of all edges of the graph representing of $\uXM$, respectively.
Moreover, $\mcE_\mcM \subseteq \psi$ where $\psi$ is the set of all edges of complete graph with vertex set $\mcV$.

\noindent {\bf Remark:} Covariance selection is presented in \cite{dempster}. Moreover, tree model as a special case for the model selection problem is discussed in subsection \ref{ssec:tree_model}.

\subsection{General detection framework}

The model selection problem is extensively studied in the literature \cite{dempster}. In many state of the art works, minimizing the KL divergence between two distributions or the maximum likelihood criterion are proposed in order to quantify the quality of the model approximation.
A different way to look at the problem of quantifying the quality of the model approximation algorithm is to formulate the problem as a detection problem \cite{lehmannHT2006}. 
Given the set of data in the detection problem, the goal is to distinguish between two hypotheses, {\it the null hypothesis} and {\it the alternative hypothesis}. 
To set up a detection problem, we need to define these two hypotheses for the model selection problem as follow
\begin{itemize}
\item[-] The null hypothesis, $\mcH_0$: The hypothesis that data is generated using the known distribution,
\item[-] The alternative hypothesis, $\mcH_1$: The hypothesis that data is generated using the model approximation distribution.
\end{itemize}

Given the set up for the null hypothesis and the alternative hypothesis, we need to define a test statistic to quantify the detection problem.
The likelihood ratio test (the Neyman-Pearson (NP) Lemma \cite{np1928}) is the most powerful test statistic where we first define the log-likelihood ratio test (LLRT) as
\begin{equation*}
l(\ux) = \textnormal{log } \frac{f_{\uX}(\ux|\mcH_1)}{f_{\uX}(\ux|\mcH_0)}
= \textnormal{log } \frac{f_{\uX_\mcM}(\ux)}{f_{\uX}(\ux)}
\end{equation*}
where $f_{\uX}(\ux|\mcH_0)$ is the distribution of random vector $\uX$ under the null hypothesis while $f_{\uX}(\ux|\mcH_1)$ is the distribution of random vector $\uX$ under the alternative hypothesis.

We then define {\it the false-alarm probability} and {\it the detection probability} by comparing the LLRT statistic under each hypothesis with a given threshold, $\uptau$,  and computing the following probabilities
\begin{itemize}
\item[-] The false-alarm probability, $P_{0}(\uptau)$, under the null hypothesis, $\mcH_0$: $P_{0}(\uptau) = \tPr ( L(\uX) \geq \uptau | \mcH_0)$, 
\item[-] The detection probability, $P_{1}(\uptau)$, under the alternative hypothesis, $\mcH_1$: $P_{1}(\uptau) = \tPr ( L(\uX) \geq \uptau | \mcH_1)$,
where random variable $L(\uX)$ is the LLRT statistic random variable. 
\end{itemize}
The most powerful test is defined by setting the false-alarm rate $P_0 (\uptau) = \bar{P_0}$ and then computing the threshold value $\uptau_0$ such that $\Pr(L(\uX) \geq \uptau_0 | \mcH_0 ) = \bar{P_0}$.

\begin{Definition}
The KL divergence between two multivariate continuous distributions $p(\uX)$ and $q(\uX)$ is defined as
$$\mcD \left( p_{\uX}(\ux)||q_{\uX}(\ux) \right) = \int_{\mcX} p_{\uX}(\ux) \log \frac{p_{\uX}(\ux)}{q_{\uX}(\ux)} \; d \ux$$
where $\mcX$ is the feasible set. \hfill \ensuremath{\blacksquare}
\end{Definition}

Throughout this paper we may use other notations such as the KL divergence between two random variable or the KL divergence between two covariance matrices for zero-mean Gaussian distribution case in order to present the KL divergence between two distributions. 

\begin{Proposition}
Expectation of the LLRT statistic under each hypothesis is
\begin{itemize}
\item[-] $\exc \left(L(\uX) | \mcH_0 \right) = - \mcD( f_{\uX}(\ux|\mcH_0) || f_{\uX}(\ux|\mcH_1 ))
= - \mcD( f_{\uX}(\ux) || f_{\uX_\mcM}(\ux))$,
\item[-] $\exc \left(L(\uX)  | \mcH_1 \right) = \mcD( f_{\uX}(\ux|\mcH_1) || f_{\uX}(\ux|\mcH_0 ))
= \mcD( f_{\uX_\mcM}(\ux) || f_{\uX}(\ux))$.
\end{itemize}
\proof
Proof is based on the KL divergence definition. \hfill \ensuremath{\blacksquare}
\end{Proposition}
\noindent {\bf Remark:} Relationship between the NP lemma and the KL divergence is previously stated in \cite{eguchi_KL_HT} with the similar straightforward calculation, where the LLRT statistic loses power when the wrong distribution is used instead of the true distribution for one of these hypotheses.

In a regular detection problem framework, the NP decision rule is to accept the hypothesis $\mcH_1$ if the LLRT statistic, $l(\ux)$, exceeds a critical value, and reject it otherwise. Moreover, the critical value is set based on the rejection probability of the hypothesis $\mcH_0$, i.e. false-alarm probability. 
Note that, we pursue a different goal in the approximation problem scenario. 
We approximate a model distribution, $f_{\uXM}(\ux)$, as close as possible to the given distribution, $f_{\uX}(\ux)$.
The closeness criterion is based on the modified detection problem framework where we compute the LLRT statistic and compare it with a threshold. In ideal case where there is no approximation error, the detection probability must be equal to the false-alarm probability for the optimal detector at all possible thresholds, i.e. the receiver operating characteristic (ROC) curve \cite{scharfSSP} that represents best detectors for all threshold values should be a line of slope $1$ passing through the origin.

In sequel, we assume that the random vector $\uX$ has zero-mean Gaussian distribution.
Thus, the covariance matrix of the random vector $\uX$ is the parameter of interest in the model selection problem, i.e. covariance selection.

\subsection{Multivariate Gaussian distribution}

Let random vector $\uX \in \mbbR^n$, have a zero-mean jointly Gaussian distribution with covariance matrix $\bSigma_{\uX}$, i.e. $\uX \sim \mcN (\uzm , \bSigma_{\uX})$ where the covariance matrix $\bSigma_{\uX}$ is positive-definite, $\bSigma_{\uX} \succ 0$. 
In this paper, the null hypothesis, $\mcH_0$, is the hypothesis that the parameter of interest is known and is equal to $\bSigma_{\uX}$ while the alternative hypothesis, $\mcH_1$, is the hypothesis that the random vector $\uX$ is replaced by the model random vector $\uXM$.
In this scenario, the model random vector $\uXM$ has a zero-mean jointly Gaussian distribution (the model approximation distribution) with covariance matrix $\bSigma_{\uXM}$, i.e. $\uXM \sim \mcN (\uzm , \bSigma_{\uXM})$ where the covariance matrix $\bSigma_{\uXM}$ is also positive-definite, $\bSigma_{\uXM} \succ 0$.
Thus, the LLRT statistic for the jointly Gaussian random vectors, $\uX$ and $\uXM$, is simplified as 
\begin{equation}
\label{eq:LLRT}
l(\ux)  = \textnormal{log } \frac{\mcN (\uzm , \bSigma_{\uXM})}{\mcN (\uzm , \bSigma_{\uX})} = - c +  k(\ux)
\end{equation}
where $c = - \frac{1}{2} \textnormal{log } (|\bSigma_{\uX} \bSigma_{\uX_{\mcM}}^{-1}|)$ is a constant and
$k(\ux) = \ux^T \bK \ux$
where $\bK = \frac{1}{2}( \bSigma_{\uX}^{-1} - \bSigma_{\uXM}^{-1})$ is an indefinite matrix with both positive and negative eigenvalues.


We define the correlation approximation matrix (CAM) associated with the covariance selection problem and dissimilarity parameters of the CAM as follows.

\begin{Definition}{\bf Correlation approximation matrix.}
	The CAM for the covariance selection problem is defined as $\bDelta \triangleq \bSigma_{\uX} \bSigma_{\uX_{\mcM}}^{-1}$ where $\bSigma_{\uX_{\mcM}}$ is the model covariance matrix. \hfill \ensuremath{\blacksquare}
\end{Definition}

\begin{Definition} {\bf Dissimilarity parameters for covariance selection problem.}
	Let $\alpha_i \triangleq \lambda_i + \lambda_i^{-1} - 2$ for $i \in \{1, \ldots,n\}$ be dissimilarity parameters of the CAM correspond to the covariance selection problem where $\lambda_i > 0$ for $i \in \{1, \ldots,n\}$ are eigenvalues of the CAM. \hfill \ensuremath{\blacksquare}
\end{Definition}

\noindent {\bf Remark:} The CAM is a positive definite matrix.
Moreover, eigenvalues of the CAM contains all information necessary to compute cost functions associated with the model selection problem.

\begin{Theorem}
\label{thm:covselect}
{\bf Covariance Selection \cite{dempster}. } Given a multivariate Gaussian distribution with covariance matrix $\bSigma_{\uX}\succ 0$, $f_{\uX}(\ux)$, and a model $\mcM$, there exists a unique approximated multivariate Gaussian distribution with covariance matrix $\bSigma_{\uXM}\succ 0$, $f_{\uXM}(\ux)$, that minimize the KL divergence, $\mcD(f_{\uX}(\ux)||f_{\uXM}(\ux))$ and satisfies the covariance selection rules, i.e. the model covariance matrix satisfies the following covariance selection rules
\begin{itemize}
\item[-] $\bSigma_{\uXM}(i,i) = \bSigma_{\uX}(i,i)$, \hspace{0.6cm}$\forall \; i \in \mcV$ 
\item[-] $\bSigma_{\uXM}(i,j) = \bSigma_{\uX}(i,j)$, \hspace{0.5cm}$\forall \; (i,j) \in \mcE_\mcM$ 
\item[-] $\bSigma_{\uXM}^{-1}(i,j) = 0$,                 \hspace{1.6cm}$\forall \; (i,j) \in \mcE_\mcM^c$ 
\end{itemize}
where the set $\mcE_\mcM^c = \psi - \mcE_\mcM$ represents the complement of the set $\mcE_\mcM$. 
\proof Proof for Gaussian distributions is given in Dempster 1972 paper \cite{dempster}. \hfill \ensuremath{\blacksquare}
\end{Theorem}

\noindent {\bf Remark:} Using the CAM definition, the constant $c$ can be written as $c = - \frac{1}{2}\log (|\bDelta|)$.
Moreover, given any covariance matrix and its model covariance matrix satisfying theorem \ref{thm:covselect}, we have $tr(\bDelta) = n$.
Thus, from the result in theorem \ref{thm:covselect} and the definition of KL divergence for jointly Gaussian distributions, we conclude $c = \mcD(f_{\uX}(\ux)||f_{\uXM}(\ux))$.

\subsection{Covariance selection example}

\begin{figure}[ht]
\begin{center}
\begin{minipage}[b]{0.4\linewidth}
	\begin{center}
		\begin{tikzpicture}[node distance = 9em, align = flush center, font = , on grid=false]	
		\tikzstyle{factor} = [ rectangle, draw, fill=green!30, text centered, minimum height=.5em ]
		\tikzstyle{state} = [ circle, draw=none, fill=white!20, text centered, minimum height=.3em ]
		\tikzstyle{fact} =   [ rounded rectangle, text centered, minimum height=.5em ]	
		
		\node[state]    					 		(F1)  {$X_1$};
		\node[state, above left		 of=F1] 	  		(F2)  {$X_2$};
		\node[state, above right	 of=F2] 			(F3)  {$X_3$};  
		\node[state,  right of=F1] 		    (F4)  {$X_4$};
		\node[fact,  right  of=F4]    (tmp) {}; 
		
		\draw	(F1) -- (F2) node[midway, font=, sloped, below] {0.9};
		\draw	(F2) -- (F3) node[midway, font=, sloped, above] {0.8};
		\draw 	(F4) -- (F3) node[midway, font=, sloped, above] {0.7};
		\draw	(F1) -- (F3) node[midway, font=, sloped, above] {0.9};
		\draw	(F2) -- (F4) node[midway, font=, sloped, above] {0.3};
		\draw 	(F4) -- (F1) node[midway, font=, sloped, above] {0.6};					
		\end{tikzpicture}
		
		(a)		
	\end{center}
\end{minipage}
\hspace{0.1\linewidth}
\begin{minipage}[b]{0.4\linewidth}
		
	\begin{center}
		\begin{tikzpicture}[node distance = 9em, align = flush center, font = , on grid=false]	
		\tikzstyle{factor} = [ rectangle, draw, fill=green!30, text centered, minimum height=.5em ]
		\tikzstyle{state} = [ circle, draw=none, fill=white!20, text centered, minimum height=.3em ]
		\tikzstyle{fact} =   [ rounded rectangle, text centered, minimum height=.5em ]	
		
		\node[state]    					 		(F1)  {$X_1$};
		\node[state, above left		 of=F1] 	  		(F2)  {$X_2$};
		\node[state, above right	 of=F2] 			(F3)  {$X_3$};  
		\node[state,  right of=F1] 		    (F4)  {$X_4$};
		\node[fact,  right  of=F4]    (tmp) {}; 
		
		\draw	(F1) -- (F2) node[midway, font=, sloped, below] {0.9};
		\draw 	(F1) -- (F3) node[midway, font=, sloped, above] {0.9};
		\draw	(F3) -- (F4) node[midway, font=, sloped, above] {0.7};
		\end{tikzpicture}	
		
		(b)
	\end{center}
\end{minipage}
\caption{(a) The complete graph; (b) The tree approximation of the complete graph.}
\label{fig:tree_example}
\end{center}
\end{figure}
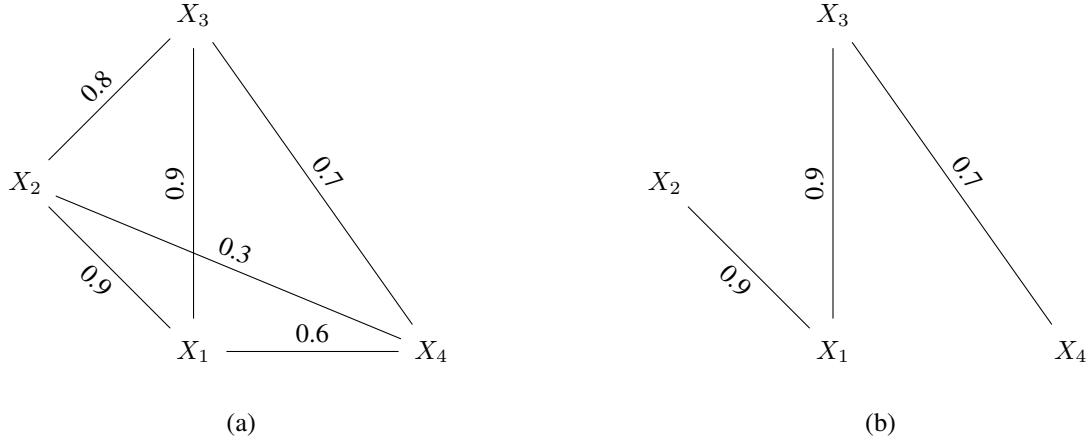

Here we choose tree approximation model as an example. Figure \ref{fig:tree_example} indicates two graphs: (a) the complete graph and (b) its tree approximation model where edges in the graph represent non-zero coefficients in the inverse of the covariance matrix \cite{dempster}.
The correlation coefficient between each pair of adjacent nodes has been written on each edge. The correlation coefficient between each pair of nonadjacent nodes is the multiplication of all correlations on the unique path that connects those nodes. 
The correlation matrix for each graph is
\begin{equation*}
\bSigma_{\uX} = 
\begin{bmatrix}
1   &   0.9   &   0.9   &   0.6   \\
0.9&    1     &   0.8   &   0.3   \\
0.9&  0.8    &     1    &   0.7    \\
0.6&  0.3    &  0.7    &     1   
\end{bmatrix}
\end{equation*}
and
\begin{equation*}
\bSigma_{\uX_{\mcT}} = 
\begin{bmatrix}
1     &   0.9   &   0.9   &   0.63     \\
0.9  &    1     &   0.81   &   0.567 \\
0.9  &  0.81    &     1    &   0.7     \\
0.63&  0.567&  0.7    &     1   
\end{bmatrix}
.
\end{equation*}

\noindent The CAM for the above example is
\begin{equation*}
\bDelta = 
\begin{bmatrix}
1    &     0    & 0.0412  & -0.0588    \\
0.0474    & 1   & 0.3042 &  -0.5098   \\
0.0474   & -0.0526   & 1 &        0   \\
0.9789  & -1.2632  &  0.1421  &  1
\end{bmatrix}
.
\end{equation*}
The CAM contains all information about the tree approximation\footnote{Dissimilarity parameters $\alpha_i$'s and eigenvalues of CAM contains all information about the tree approximation.}.
Here we assume cases that Gaussian random variables have finite, nonzero variances.

\noindent{\bf Remark:} Without loss of generality, throughout this paper we are working with normalized correlation matrices, i.e. the diagonal elements of the correlation matrices are normalized to be equal to one.

\subsection{Distribution of the LLRT statistic}

The random vector $\uX$ has Gaussian distribution under both hypotheses $\mcH_{0}$ and $\mcH_{1}$. 
Thus under both hypotheses, the real random variable, $K(\uX) = \uX^T \bK \uX$ has a generalized chi-squared distribution, i.e. the random variable, $K(\uX)$, is equal to a weighted sum of chi-squared random variables with both positive and negative weights under both hypotheses.
Let us define $\uW = \bSigma_{\uX}^{-\frac{1}{2}} \uX$ under $\mcH_{0}$ and $\uZ = \bSigma_{\uXM}^{-\frac{1}{2}} \uX$ under $\mcH_{1}$, where $\bSigma_{\uX}^{\frac{1}{2}}$ and $\bSigma_{\uXM}^{\frac{1}{2}}$ are the square root of covariance matrices $\bSigma_{\uX}$ and $\bSigma_{\uXM}$, respectively. 
Then let random vectors $\uW \sim \mcN (\uzm , \bI)$ and  $\uZ \sim \mcN (\uzm , \bI)$ have zero-mean Gaussian distributions with the same covariance matrices, $\bI$, where $\bI$ is the identity matrix of dimension $n$. 
Note that, the CAM is a positive definite matrix with $\lambda_i >0$ where $1\leq i \leq n$.
Thus, the random variable $K(\uX)$, under both hypotheses $\mcH_0$ and  $\mcH_1$ can be written as:
$$ K_0(\uX) \triangleq K(\uX) | \mcH_0 = \frac{1}{2} \sum_{i=1}^{n} (1 - \lambda_i) W_i^2$$
and
$$K_1(\uX) \triangleq K(\uX) | \mcH_1 = \frac{1}{2} \sum_{i=1}^{n} (\lambda_i^{-1} - 1) Z_i^2$$
respectively, where random variables $W_i$ and $Z_i$, are the $i$-th element of random vectors $\uW$ and $\uZ$, respectively. 
Moreover, random variables $W_i^2$ and $Z_i^2$, follow the first order central chi-squared distribution.
Note that, similarly random variable $L(\uX) \triangleq - c + K(\uX)$ is defined under each hypothesis as
$$ L_0 \triangleq L(\uX) | \mcH_0 = - c + K_0(\uX)$$
and
$$L_1 \triangleq L(\uX) | \mcH_1 = - c + K_1(\uX).$$

\noindent {\bf Remark:} As a simple consequence of the covariance selection theorem, the summation of weights for the generalized chi-squared random variable, the expectation of $K(\uX)$, is zero under the hypothesis $\mcH_0$, i.e. $\exc (K_0(\uX))  = \frac{1}{2} \sum_{i=1}^{n} (1 - \lambda_i) = 0$ \cite{dempster}, and this summation is positive under the hypothesis  $\mcH_1$, i.e. $\exc (K_1(\uX))  = \frac{1}{2} \sum_{i=1}^{n} (\lambda_i^{-1} - 1) \geq 0$.

\section{The ROC Curve and the AUC Computation}
\label{sec:AUC}

\subsection{The receiver operating characteristic curve}

The receiver operating characteristic (ROC) curve is the parametric curve where the detection probability is plotted versus the false-alarm probability for all thresholds, i.e. each point on the ROC curve represents a pair of $( P_0(\uptau) , P_1(\uptau) )$ for a given threshold $\uptau$.
Set $z = P_0(\uptau)$ and $\eta = P_1(\uptau)$, the ROC curve is $\eta = h(z) $.
If $P_0(\uptau)$ has an inverse function, then the ROC curve is $h(z)=P_1(P_0^{-1}(z))$.
In general, the ROC curve, $h(z)$, has the following properties \cite{scharfSSP}
\begin{itemize}
\item[-] $h(z)$ is concave and increasing,
\item[-] $h'(z)$ is positive and decreasing,
\item[-] $\int_{0}^{1} h'(z) \, dz\leq 1$. 
\end{itemize}
Note that, for the ROC curve, the slope of the tangent line at a given threshold, $h'(z)$, gives the likelihood ratio for the value of the test.

\noindent{\bf Remark:} For the ROC curve for our Gaussian random vectors we have $h'(z)$ is positive, continuous and decreasing in interval $[0,1]$ with right continuity at $0$ and left continuity at $1$. Moreover, 
$$\int_{0}^{1} h'(z) \, dz = 1$$ 
since  $h(0)=0$ and $h(1)=1$.

\begin{figure}[ht]
\begin{centering}
\includegraphics[width=.9\linewidth]{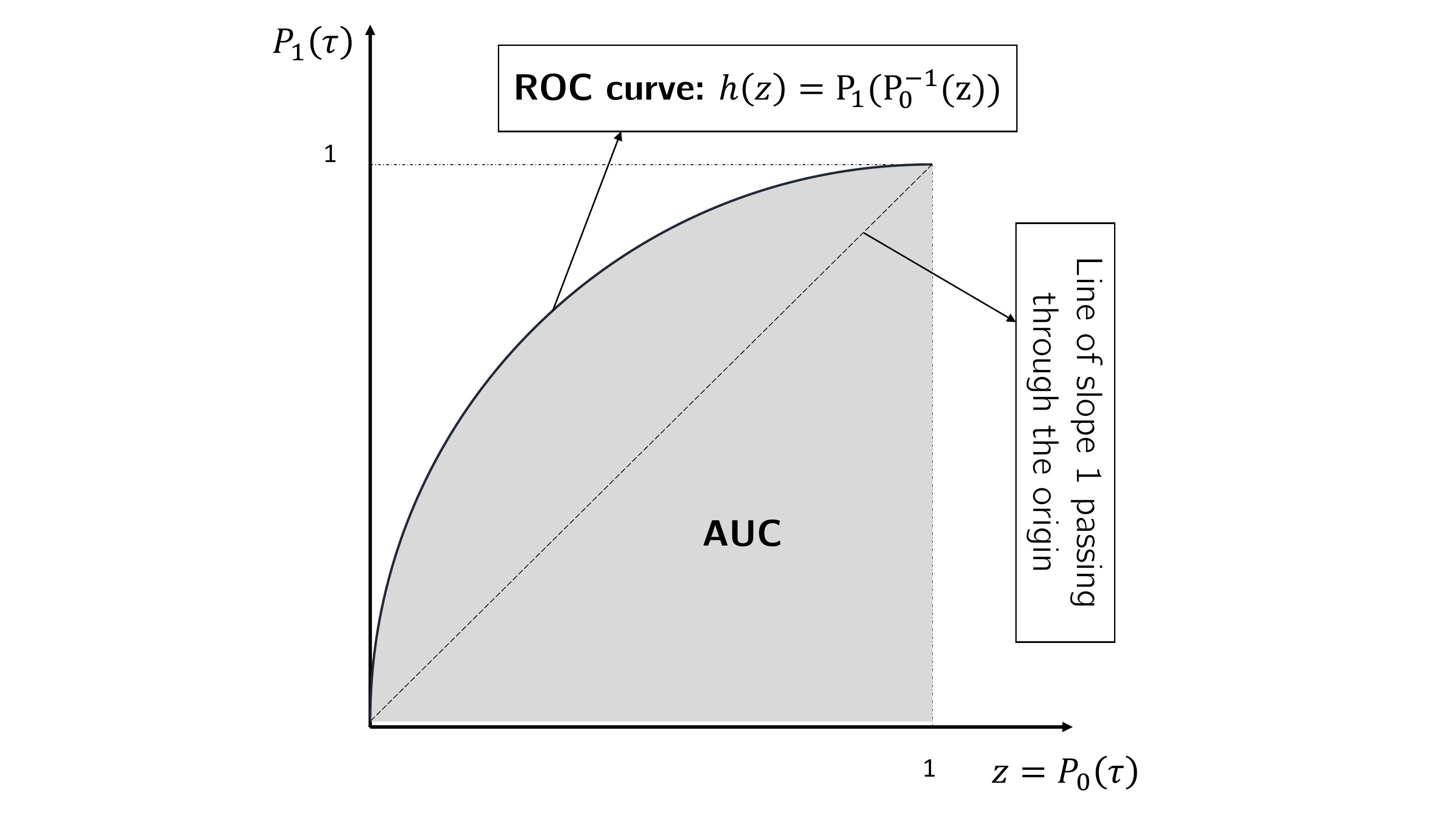}
\caption{The ROC curve and the area under the ROC curve. Each point on the ROC curve indicates a detector with given detection and false-alarm probabilities.}
\label{fig:ROCcurve}
\end{centering}
\end{figure}

\begin{Definition}
	Let $f_{L_0}(l)$ and $f_{L_1}(l)$ be the probability density function (PDF) of the random variables $L_0$ and $L_1$, respectively.
	\hfill \ensuremath{\blacksquare}
\end{Definition}

\begin{Lemma}
\label{lem:RKL}
Given the ROC curve, $h(z)$, we can compute following KL divergences
\begin{equation*}
\mcD \left(f_{L_1}(l) || f_{L_0}(l)\right) = - \int_{0}^{1} \log \left( h'(z) \right) \, dz.
\end{equation*}
and
\begin{align*}
\mcD \left(f_{L_0}(l) || f_{L_1}(l)\right) & = - \int_{0}^{1}  h'(z) \, \log \left( h'(z) \right) \, dz  \\
&\stackrel{(*)}{=} - \int_{0}^{1} \log \left( \frac{d \, h^{-1}(\eta)}{d\,\eta}\right) \, d\eta 
\end{align*}
where (*) holds if the ROC curve, $\eta = h(z)$, has an inverse function.
\proof These results are consequence of the Radon-Nikod\'{y}m theorem \cite{RNT}. Simple, alternative calculus based proofs are given in appendix \ref{apx:RKL}. \hfill \ensuremath{\blacksquare}
\end{Lemma}

\subsection{Area under the curve}

As discussed previously we examine the ROC with a goal that the model approximation results with the ROC being a line of slope 1 passing through the origin. This is in contrast to the conventional detection problem where we want to distinguish between the two hypotheses and ideally have an ROC that is a unit step function.
Area under the curve (AUC) is defined as the integral of the ROC curve (figure \ref{fig:ROCcurve}) and is a measure of accuracy in decision problems.

\begin{Definition}
The area under the ROC cure (AUC) is defined as 
\begin{equation}
\label{eq:AUCdef}
AUC = \int_{0}^{1} h(z)\,  d\, z = \int_{0}^{1} P_1(\uptau) \, d P_0(\uptau)
\end{equation}
where $\uptau$ is the detection problem threshold.\hfill \ensuremath{\blacksquare}
\end{Definition}

\noindent {\bf Remark:} The AUC is a measure of accuracy for the detection problem and $\sfrac{1}{2} \le {\rm AUC} \le 1$. Note that, in conventional decision problems, the AUC is desired to be as close as possible to $1$ while in approximation problem presented here we want the AUC to be close to $\sfrac{1}{2}$.

\begin{Theorem}{\bf Statistical property of AUC \cite{hanleyAUC}.}
\label{thm:AUC}
The AUC for the LLRT statistic is 
\begin{equation*}
AUC = \textnormal{Pr} \left( L_1 >  L_0  \right).
\end{equation*}
\hfill\ensuremath{\blacksquare}
\end{Theorem}

\begin{Corollary}
\label{lem:AUC}
From theorem \ref{thm:AUC}, when PDFs for the LLRT statistic under both hypotheses exist, we can compute the AUC as
\begin{equation}
\label{eq:AUCdef2}
AUC = \int_{0}^{\infty} f_{L_1}(l) \, \star \, f_{L_0}(l) \,dl
\end{equation}
where $ f_{L_1}(l) \, \star \, f_{L_0}(l) \triangleq \int_{-\infty}^{\infty} f_{L_1}(\uptau) \, f_{L_0}(l+\uptau) \, dl$ is the cross-correlation between $f_{L_1}(l)$ and $f_{L_0}(l)$.
\proof A proof based on the definition of the AUC \eqref{eq:AUCdef}, is given in \cite{ITA2016}. \hfill \ensuremath{\blacksquare}
\end{Corollary}

Let us define the difference LLRT statistic random variable as $L_{\Delta} =  L_1 - L_0$. Then, we get
\begin{align*}
AUC & =  \textnormal{Pr} \left( L_{\Delta}>0 \right) \\
&= 1 - F_{L_{\Delta}} (0)
\end{align*}
where $F_{L_{\Delta}} (l)$ is the cumulative distribution function (CDF) for random variable $L_{\Delta}$.

The two conditional random variables $ L_0$ and $L_1$ are independent\footnote{By the definition in the detection problem.}.
Thus, the cross-correlation between the corresponding two distributions is the distribution of the difference LLRT statistic, $L_{\Delta}$.
We can write the random variable $L_{\Delta}$ as
\begin{align*}
L_{\Delta} &= -c + K_1(\uX)  - \left( -c + K_0(\uX) \right) \\
& = K_1(\uX)  -  K_0(\uX).
\end{align*}
Replacing the definition for $K_0(\uX)$ and $K_1(\uX) $, we have
$$ L_{\Delta} = \frac{1}{2} \sum_{i=1}^{n} (\lambda_i^{-1} - 1) Z_i^2 - \frac{1}{2} \sum_{i=1}^{n} (1 - \lambda_i) W_i^2 . $$
We can rewrite the difference LLRT statistic, $L_{\Delta}$, in an indefinite quadratic form as
\begin{equation*}
L_{\Delta} = \frac{1}{2} \uV^T ( \bLambda - \bI ) \uV
\end{equation*}
where 
$$\uV = 
\begin{bmatrix}
\uW \\\\
 \uZ
\end{bmatrix} 
$$
and 
$$\bLambda =
\begin{bmatrix}
\lambda_1 &  &                     &   &  &  \\
                 & \ddots  &                    &   &  {\bf 0} &  \\
                 &   & \lambda_n   &  &  &  \\
                 &   &   &   \lambda^{-1}_1  &   & \\
                 &  {\bf 0} &   &   &  \ddots & \\
                 &   &   &   &  & \lambda^{-1}_n   \\
\end{bmatrix}
.
$$

\subsection{Generalized Asymmetric Laplace distribution}
The difference LLRT statistic random variable, $L_{\Delta}$, follows the generalized asymmetric Laplace (GAL) distribution\footnote{Also known as the variance-gamma distribution or the Bessel function distribution.} \cite{VGdist}. 
For a given $i$ where $i \in \{1, \ldots,n\}$, we define random variable $L_{\Delta_i}$ as
\begin{equation}
\label{eq:glrv}
L_{\Delta_i} = \frac{ \lambda_i -1 }{2} W_i^2  -  \frac{ 1 - \lambda_i^{-1} }{2} Z_i^2 .
\end{equation}
Then, difference LLRT statistic random variable, $L_{\Delta}$, can be written as
$$ L_{\Delta}=   \sum_{i=1}^{n} L_{\Delta_i}$$
where $L_{\Delta_i}$'s are independent and have GAL distributions at position $0$ with mean $\sfrac{\alpha_i}{2}$ and PDF \cite{VGdist}
\begin{equation}
\label{eq:VGdist}
f_{L_{\Delta_i}} (l) = \frac{e^\frac{l}{2}}{\pi \sqrt{\alpha_i}} \,K_0 \left( \sqrt{\alpha_i^{-1} + \frac{1}{4}}\;|l| \right), \quad\quad l \neq 0
\end{equation}
where $ K_0 (-)$ is the modified Bessel function of second kind \cite{MF_Bessel_K}.
The moment generating function (MGF) for this distribution is
$$M_{L_{\Delta_i}} (t) = \frac{1}{\sqrt{1 - \alpha_i t - \alpha_i t^2}}$$
for all $t$'s that satisfies $1 - \alpha_i t - \alpha_i t^2 > 0$.
From \eqref{eq:glrv}, the MGF derivation for the GAL distribution is straightforward and is the multiplication of two MGFs for the chi-squared distribution.

The distribution of the difference LLRT statistic random variable, $L_{\Delta}$, is
$$f_{L_{\Delta}}(l) = \bigast_{i=1}^{n} f_{L_{\Delta_i}} (l) $$
where $\bigast_{i=1}^{n}$ is the notation we use for convolution of $n$ functions together.
Note that, although the distribution of random variables $L_{\Delta_i}$'s in \eqref{eq:VGdist} has discontinuity at $l = 0$, the distribution of random variable $L_{\Delta}$ is continuous if there are at least two distribution with non-zero parameters, $\alpha_i$'s, in the aforementioned convolution.
Moreover, the MGF for $f_{L_{\Delta}}(l)$ can be computed by multiplying MGFs for $L_{\Delta_i}$ as
\begin{equation}
\label{eq:glMGF}
M_{L_{\Delta}} (t) = \prod_{i=1}^{n} M_{L_{\Delta_i}} (t)
\end{equation}
for all $t$'s in the intersection of all domains of $M_{L_{\Delta_i}} (t)$.
The smallest of such intersections is $-1<t<0$.

\subsection{Analytical expression for AUC}
To compute the CDF of random variable $L_{\Delta}$, we need to evaluate a multi-dimensional integral of jointly Gaussian distributions \cite{provost_exact} or we need to approximate this CDF \cite{ha_provost_approx}. 
More efficiently, as discussed in \cite{indefiniteH} for the real valued case,  the CDF of the random variable $L_{\Delta}$ can be expressed as a single-dimensional integral of a complex function\footnote{This is the transform to the frequency domain for an arbitrary $\beta$.} in the following form
$$F_{L_{\Delta}}( l) = \frac{1}{2 \pi } \int_{-\infty}^{\infty} \frac{e^{ \frac{l}{2} (j \omega + \beta)}}{j \omega + \beta}
\frac{1}{\sqrt{|\bI + \textcolor{black}{\frac{1}{2}}(\bLambda - \bI)(j \omega + \beta)|}} \; d \omega$$
where $\beta>0$ is chosen such that matrix $\bI + \frac{\beta}{2}(\bLambda - \bI)$, is positive definite and simplifies the evaluation of the multivariate Gaussian integral \cite{indefiniteH}.

\noindent {\bf Special case:} When $\bLambda = \bI$, i.e. the given covariance obeys the model structure, then
$$AUC = 1 - F_{L_{\Delta}}( 0) = 1 - \frac{1}{2 \pi } \int_{-\infty}^{\infty} \frac{1}{j \omega + \beta} \; = \frac{1}{2}$$
for $\beta>0$ and is also independent of the value of the parameter $\beta$.

Picking an appropriate value for the parameter $\beta$ \footnote{The parameter $\beta$ is picked such that $\bI + \frac{\beta}{2}(\bLambda - \bI) \succ 0$ and $\beta =2$ always satisfies this condition since $\bLambda \succ 0$.}, the AUC can be numerically computed by evaluating the following one dimension complex integral
$$ AUC = 1 -  \frac{1}{2 \pi } \int_{-\infty}^{\infty} \frac{1}{j \omega + \beta}
\frac{1}{\sqrt{|\bI + \textcolor{black}{\frac{1}{2}}(\bLambda - \bI)(j \omega + \beta)|}} \; d \omega . $$
Furthermore, since $\bLambda \succ 0$, choosing $\beta = 2$ and changing variable as $\nu = \sfrac{\omega}{2}$, we have
\begin{equation}
\label{eq:AUCeq}
AUC= 1 - \frac{1}{2 \pi } \int_{-\infty}^{\infty} \frac{1}{j \nu + 1}
\frac{1}{\sqrt{|\bLambda + j \nu (\bLambda - \bI)|}} \; d \nu .
\end{equation}
Moreover, $|\bLambda + j \nu (\bLambda - \bI)| = \prod_{i=1}^{p} ( 1 + \alpha_i \nu^2 - j \alpha_i \nu )$. This equation shows that the AUC only depends on $\alpha_i$'s.

\noindent {\bf Remark:} Since the AUC integral in \eqref{eq:AUCeq} can not be evaluated in closed form, it can not be used directly in obtaining model selection algorithms. Numerical evaluation of the AUC using the one dimensional complex integral \eqref{eq:AUCeq} is very efficient and fast comparing to numerical evaluation of a multi-dimensional integral of jointly Gaussian CDF.


\section{Analytical Bounds for the AUC}
\label{sec:bound}

As in the previous section we present an analytical expression for the AUC, in this section, we presents analytical lower and upper bounds for the AUC. These bounds will give us incite on how the AUC behave.

\subsection{Lower bound for the AUC (Chernoff bound application)}

Given the MGF for the difference LLRT statistic distribution \eqref{eq:glMGF}, we can apply the Chernoff bound \cite{cover} to find a lower bound for the AUC or upper bound for the CDF of the difference LLRT statistic random variable, $L_{\Delta}$, evaluated at zero).
\begin{Proposition}
Lower bound for the AUC is
\begin{equation}
\tPr \left( L_{\Delta}>0 \right) \geq \max \left\{ \frac{1}{2} ,  1 - e^{ - \frac{1}{2} \sum_{i=1}^{n} \log\left(1+\frac{\alpha_i}{4}\right) } \right\}
\end{equation}
\proof
One-half is a trivial lower bound for AUC.
To achieve a non-trivial lower bound, we apply Chernoff bound \cite{cover} as follow
$$\textnormal{Pr} \left( L_{\Delta}<0 \right) \leq \inf_{t}  \; M_{L_{\Delta}} (t).$$
To complete the proof we need to solve the right-hand-side (RHS) optimization problem.
\\
{\bf Step 1:} First derivatives of $M_{L_{\Delta}} (t)$ is
\begin{align*}
\frac{d}{d \,t} M_{L_{\Delta}} (t) & = M_{L_{\Delta}} (t) \\
& \left( \frac{1}{2} \sum_{i=1}^{n} \frac{\lambda_i-1}{1 - (\lambda_i-1) t} + \frac{\lambda_i^{-1}-1}{1 - (\lambda_i^{-1}-1) t}   \right) \\
 & = M_{L_{\Delta}} (t) (1+2 t) \sum_{i=1}^{n} \frac{\alpha_i}{2(1 - \alpha_i t - \alpha_i t^2)}.
\end{align*}
Clearly, first derivative is zero for $t = -\sfrac{1}{2}$ which is in the feasible domain of the MGF for the difference LLRT statistic.
Note that, the smallest feasible domain is $-1<t<0$.
\\
{\bf Step 2:} Second derivatives of $M_{L_{\Delta}} (t)$ is
\begin{align*}
& \frac{d^2}{d \,t^2} M_{L_{\Delta}} (t)  =  \\
& M_{L_{\Delta}} (t) \left( \frac{1}{4} \sum_{i=1}^{n} \frac{\lambda_i-1}{1 - (\lambda_i-1) t} + \frac{\lambda_i^{-1}-1}{1 - (\lambda_i^{-1}-1) t}   \right)^2 +\\
& M_{L_{\Delta}} (t) \left( \frac{1}{4} \sum_{i=1}^{n} \frac{(\lambda_i-1)^2}{(1 - (\lambda_i-1) t)^2} + \frac{(\lambda_i^{-1}-1)^2}{(1 - (\lambda_i^{-1}-1) t)^2}   \right).
\end{align*}
Therefore, we conclude that the second derivative is positive and thus the optimal solution to the RHS optimization problem is at $t = -\frac{1}{2}$.
Replacing that in the definition of the moment generation function which results in the following bound
\begin{equation*}
\tPr \left( L_{\Delta}\leq0 \right) <  \prod_{i=1}^{n} \frac{2}{\sqrt{4+\alpha_i}}
\end{equation*}
which can be written as
\begin{equation*}
\tPr \left( L_{\Delta}>0 \right) \geq  1 - \prod_{i=1}^{n} \frac{2}{\sqrt{4+\alpha_i}}
\end{equation*}
which completes the proof.\hfill \ensuremath{\blacksquare}
\end{Proposition}

\subsection{Upper Bound for the AUC}

In this section, we present a parametric upper bound for the AUC, but first, we need to present the following results.


\begin{Lemma} 
\label{lem:IPKL}
{\bf Invariance property of the KL divergence for the LLRT statistic.} We have
$$\mcD(f_{L_1}(l) || f_{L_0}(l)) \leq \mcD( f_{\uX}(\ux|\mcH_1) || f_{\uX}(\ux|\mcH_0 ))$$
and
$$\mcD(f_{L_0}(l) || f_{L_1}(l)) \leq \mcD( f_{\uX}(\ux|\mcH_0) || f_{\uX}(\ux|\mcH_1 )).$$
\proof This lemma is an special case of the invariance property of the KL divergence \cite{KL_book}. By picking appropriate measurable mapping, here appropriate quadratic function for each equation of the above equations, we conclude the lemma.   \hfill \ensuremath{\blacksquare}
\end{Lemma}

\begin{figure}[ht]
\begin{centering}
\includegraphics[width=1\linewidth]{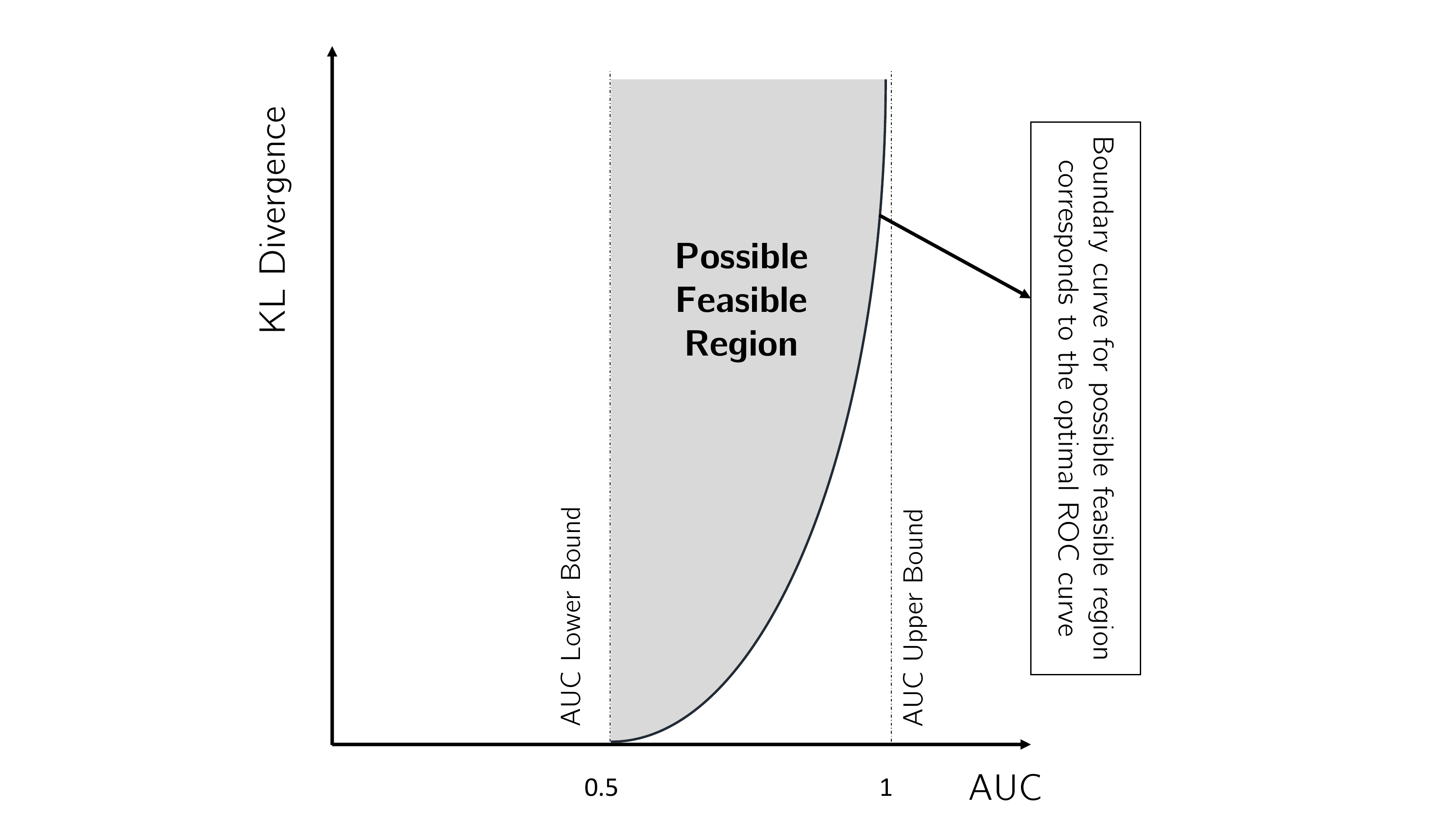}
\caption{Possible feasible region for the AUC and the Kl divergence pair for all possible detectors or equivalently all possible ROC curves (the KL divergence is between the LLRT statistics under different hypotheses, i.e. $\mcD(f_{L_0}(l) || f_{L_1}(l))$ or $\mcD(f_{L_1}(l) || f_{L_0}(l))$.)}
\label{fig:PFR}
\end{centering}
\end{figure}

\begin{Definition}
{\bf Feasible Region.} The AUC and the KL divergence pair is lying in the feasible region (figure \ref{fig:PFR}) for all possible detectors (ROC curves), i.e. no detector with the AUC and the KL divergence pair lie outside the feasible region\footnote{The definition of the feasible region here is inspired by the joint range of f- divergences \cite{fD_JP_peter}.}. 
\end{Definition}

\begin{Theorem}
\label{thm:ART}
{\bf Possible feasible region for the AUC and the KL divergence.} Given the ROC curve, the parametric possible feasible region as shown in figure \ref{fig:PFR} can be expressed using the positive parameter $a > 0$ as
\begin{equation*}
\textnormal{Pr} \left( L_{\Delta}>0 \right) = \frac{1}{1-e^{-a}} - \frac{1}{a}
\end{equation*}
and
\begin{equation*}
\mcD_{l}^{*} \geq \log(a) +  \frac{a}{e^{a} - 1} - 1 - \log(1-e^{-a})
\end{equation*}
where
\begin{equation*}
\mcD_{l}^{*} = \min \left\{ \mcD(f_{L_1}(l) || f_{L_0}(l)) \, ,\,  \mcD(f_{L_0}(l) || f_{L_1}(l)) \right\}.
\end{equation*}
\proof Proof is given in the appendix \ref{apx:LMS}.  \hfill \ensuremath{\blacksquare}
\end{Theorem}

Theorem \ref{thm:ART} formulates the relationship between the AUC and the KL divergence. {\it The results of this theorem is generally true for any LLRT statistic}. Theorem \ref{thm:ART} states that for any valid ROC corresponds to a detector, the pair of AUC and KL divergence {\it must} lie in the possible feasible region (figure \ref{fig:PFR}), i.e. outside of this region is infeasible.
This possible feasible region results in the general upper bound for AUC.

Since computing the distribution of the LLRT statistics is not straight forward in most cases, proposition \ref{prop:ubAUC}, relaxes the Theorem \ref{thm:ART} by bounding the KL divergence between the LLRT statistics using the the invariance property of KL divergence for the LLRT statistic (lemma \ref{lem:IPKL}).
\begin{Proposition}
\label{prop:ubAUC}
The parametric upper bound for AUC is
\begin{equation*}
\textnormal{Pr} \left( L_{\Delta}>0 \right) = \frac{1}{1-e^{-a}} - \frac{1}{a}
\end{equation*}
and
\begin{equation*}
\mcD^{*} \geq \log(a) +  \frac{a}{e^{a} - 1} - 1 - \log(1-e^{-a})
\end{equation*}
where $a>0$ is a positive parameter and
\begin{align}
\label{eq:Dmin}
\mcD^{*} = \min \{ \; & \mcD( f_{\uX}(\ux|\mcH_1) || f_{\uX}(\ux|\mcH_0 )) \, ,\, \\\notag & \mcD( f_{\uX}(\ux|\mcH_0) || f_{\uX}(\ux|\mcH_1 )) \; \}.
\end{align}
\proof Proof is based on the lemma \ref{lem:IPKL} and the possible feasible region presented in the theorem \ref{thm:ART}.
From the lemma \ref{lem:IPKL}, we have  
$$\mcD_{l}^{*} \leq \mcD^{*}.$$
Then, using the result in the theorem \ref{thm:ART}, we get the parametric upper bound.
\hfill \ensuremath{\blacksquare}
\end{Proposition}

\subsection{Asymptotic behavior for AUC bounds}

\begin{Proposition}{\bf Asymptotic behavior of the lower bound.}
We have
\begin{equation*}
\textnormal{Pr} \left( L_{\Delta}>0 \right) \geq 1 - e^{-n\left(1-\frac{1}{n}\sum_{i=1}^{n}(1+\frac{\alpha_i}{8})^{-1}\right)}.
\end{equation*}
\proof Applying the inequality
\begin{align*}
\frac{2x}{2+x} < \log (1+x)
\end{align*}
for $x>0$, we achieve the result.
\hfill \ensuremath{\blacksquare}
\end{Proposition}

\begin{Proposition}{\bf Asymptotic behavior of the upper bound.}
The parametric upper bound for AUC has the following asymptotic behavior
\begin{equation*}
\label{prop:asyUB}
\textnormal{Pr} \left( L_{\Delta}>0 \right) \leq 1 - e^{-\mcD^{*}-1}
\end{equation*}
where $\mcD^{*} $ is given in \eqref{eq:Dmin}.
\proof Proof is as follows.
\begin{align*}
- \log \left( 1 - \textnormal{Pr} \left( L_{\Delta}>0 \right) \right) & = - \log \left(  \frac{1}{e^{a}-1} + \frac{1}{a} \right) \\
& \leq \log \left( a \right) \\
& \leq \mcD^{*}+1.
\end{align*}
Applying the exponential function to both sides of the above inequality we conclude the upper bound.
\hfill \ensuremath{\blacksquare}
\end{Proposition}

\begin{figure}[ht]
\begin{centering}
\includegraphics[width=.8\linewidth]{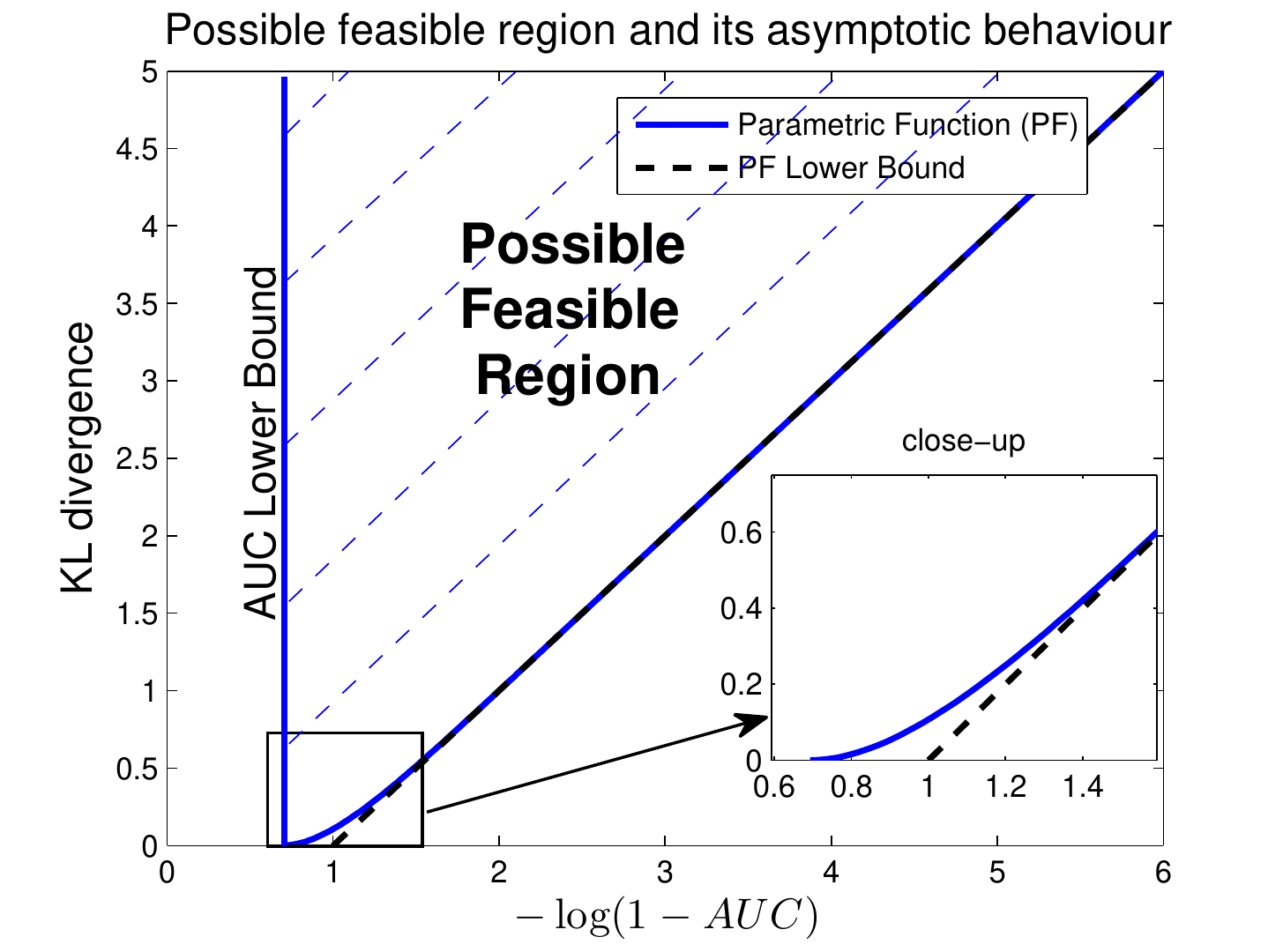}
\caption{Log-scale of the possible feasible region and its asymptotic behavior (linear line) for the AUC and the KL divergence pair for all possible detectors or equivalently all possible ROC curves (the KL divergence is between the LLRT statistics under different hypotheses, i.e. $\mcD(f_{L_1}(l) || f_{L_0}(l))$ or $\mcD(f_{L_0}(l) || f_{L_1}(l))$.) Close-up part shows the non-linear behavior of the possible feasible region around one.}
\label{fig:FRLS}
\end{centering}
\end{figure}

\begin{figure}[ht]
\begin{centering}
\includegraphics[width=.8\linewidth]{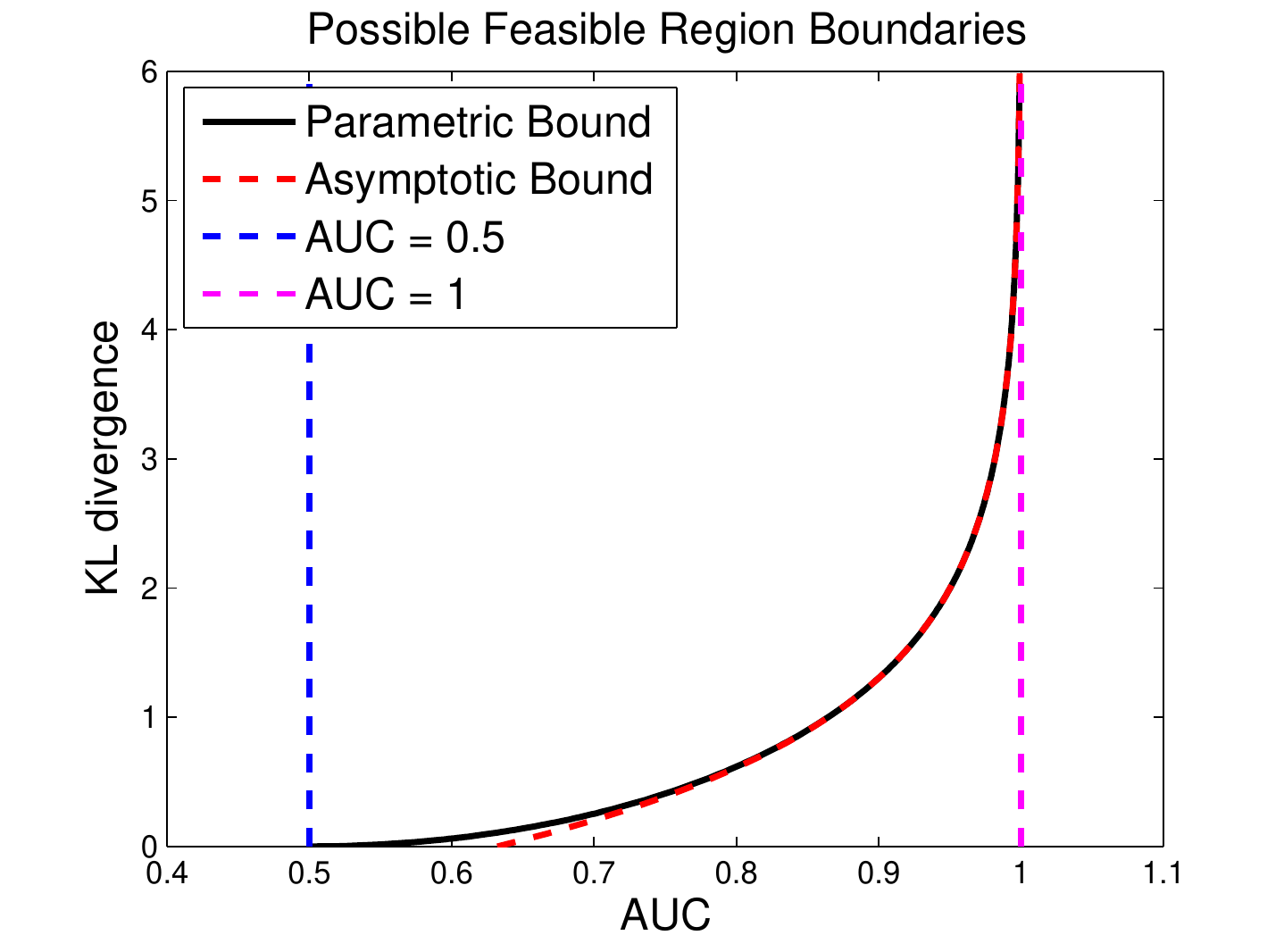}
\caption{The possible feasible region boundaries and its asymptotic behavior for the AUC and the KL divergence pair for all possible detectors or equivalently all possible ROC curves (the KL divergence is between the LLRT statistics under different hypotheses, i.e. $\mcD(f_{L_0}(l) || f_{L_1}(l))$ or $\mcD(f_{L_1}(l) || f_{L_0}(l))$.).}
\label{fig:FRRS}
\end{centering}
\end{figure}

Figure \ref{fig:FRLS} shows the possible feasible region and the asymptotic behavior log-scale. As it is shown in this figure, the parametric upper bound can be approximated with a straight line specially for large values of the parameter $a$ (the result in proposition \ref{prop:asyUB}).
Also, figure \ref{fig:FRRS} shows the possible feasible region and the asymptotic behavior in regular-scale.

\section{Examples and Simulation Results}
\label{sec:examples}

In this section, we consider some examples of covariance matrices for Gaussian random vector $\uX$. We pick the tree structure as the graphical model corresponds to the covariance selection problem.
In our simulations, we compare the numerically evaluated AUC and its lower and upper bounds and discuss their asymptotic behavior as the dimension of the graphical model, $n$, increases.

\subsection{Tree approximation model}
\label{ssec:tree_model}

The maximum order of the lower order distributions in tree approximation problem is two, i.e. no more than pairs of variables.
Let $\uX_{\mcT} \sim \mcN (\uzm , \bSigma_{\uX_{\mcT}})$ have the graph representation $\mcG_{\mcT}=(\mcV, \mcE_{\mcT})$ where $\mcE_{\mcT} \subseteq \psi$ is a set of edges that represents a tree structure.
Let $\uX_r \sim \mcN (\uzm , \bSigma_{\uX_r})$ have the graph representation $\mcG_r=(\mcV, \mcE_r)$ where $\mcE_r \subseteq \mcE_{\mcT}$ is the set of all edges in the graph of $\uX_r$.
The joint PDF for elements of random vector $\uX_r$ can be represented by joint PDFs of two variables and marginal PDFs in the following convenient form
\begin{equation}
\label{eq:prod_approx}
f_{\uX_r}(\ux_r) = \prod_{(u,v) \in \mcE_r} \frac{f_{\uX^u,\uX^v}(\ux^u , \ux^v) }{f_{\uX^u}(\ux^u) f_{\uX^v}(\ux^v)} \prod_{u \in \mcV} f_{\uX^u}(\ux^u).
\end{equation}
Using equation \eqref{eq:prod_approx} we can then easily construct a tree using iterative algorithms (such as the Chow-Liu algorithm \cite{chowliu} combined with the Kruskal \cite{kruskal} algorithm or the Prim \cite{prim} algorithm) by adding edges one at a time \cite{Lbanded_kavcic}.
Consider the sequence of random vectors $\uX_r$ with $0\leq r \leq|\mcE_{\mcT}|$, where $\uX_r$ is recursively generated by augmenting a new edge, $(i,j) \in \mcE_r$, to the graph representation of $\uX_{r-1}$.
For the special case of Gaussian distributions, $\bSigma_{\uX_r}$ has the following recursive formulation \cite{Lbanded_kavcic}
\begin{equation*}
\label{eq:SigXl}
\bSigma_{\uX_r}^{-1} = \bSigma_{\uX_{r-1}}^{-1} + \bSigma_{i,j}^{\dagger} - \bSigma_{i}^{\dagger} - \bSigma_{j}^{\dagger} \; , \quad \forall \; 0\leq r \leq|\mcE_{\mcT}|
\end{equation*}
where
$\bSigma_{i,j}^{\dagger} = [\ue_i \; \ue_j] \bSigma_{i,j}^{-1} [\ue_i \; \ue_j]^T $
and
$\bSigma_{i}^{\dagger}   = \ue_i \bSigma_{i}^{-1} \ue_i^T$
where $\ue_i$ is a unitary vector with $1$ at the $i$-th place and $\bSigma_{i,j}$ and $\bSigma_{i}$ are the $2$-by-$2$ and $1$-by-$1$ principle sub-matrices of $\bSigma_{\uX}$,
with initial step $\bSigma_{\uX_0} \!\!= \! diag(\bSigma_{\uX})$ where $diag(\bSigma_{\uX})$ represents a diagonal matrix with diagonal elements of $\bSigma_{\uX}$.

\noindent {\bf Remark:} 
For all $0\leq r \leq|\mcE_{\mcT}|$, we have
\begin{enumerate}
\item $\textnormal{tr}(\bSigma_{\uX_r}) = \textnormal{tr}(\bSigma_{\uX}) $
\item $\textnormal{tr}(\bSigma_{\uX}\bSigma_{\uX_r}^{-1}) = n.$
\item $\mcD ( f_{\uX}(\ux) || f_{\uX_r}(\ux) ) = -\frac{1}{2}\textnormal{log}( |\bSigma_{\uX}\bSigma_{\uX_r}^{-1} |)$
\item $|\bSigma_{\uX}| \leq \ldots \leq |\bSigma_{\uX_r}| \leq \ldots \leq |\bSigma_{\uX_0}| = |diag(\bSigma_{\uX})|$
\item $H(\uX) \leq \ldots \leq H(\uX_r) \leq \ldots \leq H(\uX_0).$
\end{enumerate}

Tree approximation models are interesting to study since there are algorithms such as Chow-Liu \cite{chowliu} combined by the Kruskal \cite{kruskal} or the Prim's \cite{prim} that efficiently compute the model covariance matrix from the graph covariance matrix.

%
%

\subsection{Toeplitz example}
Here, we assume that the covariance matrix $ \bSigma_{\uX}$ has a Toeplitz structure with ones on the diagonal elements and the correlation coefficient $\rho > - \frac{1}{(n-1)}$ as off diagonal elements
\begin{equation*}
 \bSigma_{\uX} =
\begin{bmatrix}
                1 &  \rho & \ldots       &\rho      \\
                \rho  & \ddots  & \ddots  &  \vdots          \\
              \vdots & \ddots &   \ddots  & \rho \\
         \rho  &   \ldots   &    \rho   &  1 \\
\end{bmatrix}
.
\end{equation*}
For the tree structure model, all possible tree structured distributions satisfying \eqref{eq:prod_approx} have the same KL divergence to the original graph, i.e. $\mcD( f_{\uX}(\ux) || f_{\uX_\mcT}(\ux))$ is constant for all possible connected tree approximation model for this example. The reason is that all the weights computed by the Chow-Liu algorithm to construct the weighted graph associated with this problem are the same and are equal to $- \frac{1}{2}\log (1+\rho^2)$, which only depends on the correlation coefficient $\rho$.
In the sequel, we test our results for two tree structured networks: a star network and a chain network.

\subsubsection{Star approximation}
The star covariance matrix is as follows (all the nodes are connected to the first node)\footnote{All $n$ possible star networks have the same performance.}
\begin{equation*}
 \bSigma_{\uX_\mcT}^{star} =
\begin{bmatrix}
                 1 & \rho  & \ldots & \ldots &  \rho     \\
             \rho & \ddots  &         \rho^2        &  \ldots  &   \rho^2    \\
             \vdots &  \rho^2   &  \ddots     &  \ddots&  \vdots \\
             \vdots & \vdots   &  \ddots     & \ddots & \rho^2   \\ 
            \rho &  \rho^2   &  \ldots &  \rho^2  &  1  \\
\end{bmatrix}
.
\end{equation*}
For this example, the KL divergence and the Jeffreys divergence can be computed in closed form as
$$\mcD (\uX||\uX_{star}) = \frac{1}{2} (n-1) \log(1+\rho) -  \frac{1}{2} \log( 1 + (n-1) \rho ) $$
and
$$\mcD_{\mcJ} (\uX \, , \, \uX_{star}) = \frac{(n-1)(n-2) \rho^2}{2(1 + (n-1) \rho)}$$
respectively, where 
$$\mcD_{\mcJ} (\uX \, , \, \uX_{star}) = \mcD (\uX||\uX_{star}) + \mcD (\uX_{star}||\uX)$$
is the Jeffreys divergence \cite{jeffreysD}. 
Moreover, for large values of $n$ we have that
$$\mcD (\uX||\uX_{star}) \approx \frac{n}{2} \log(1+\rho) $$
and
$$\mcD_{\mcJ} (\uX \, , \, \uX_{star}) \approx \frac{n}{2} \rho .$$
%

\begin{figure}[ht]
\begin{minipage}[b]{0.48\linewidth}
\includegraphics[width=1.08\linewidth]{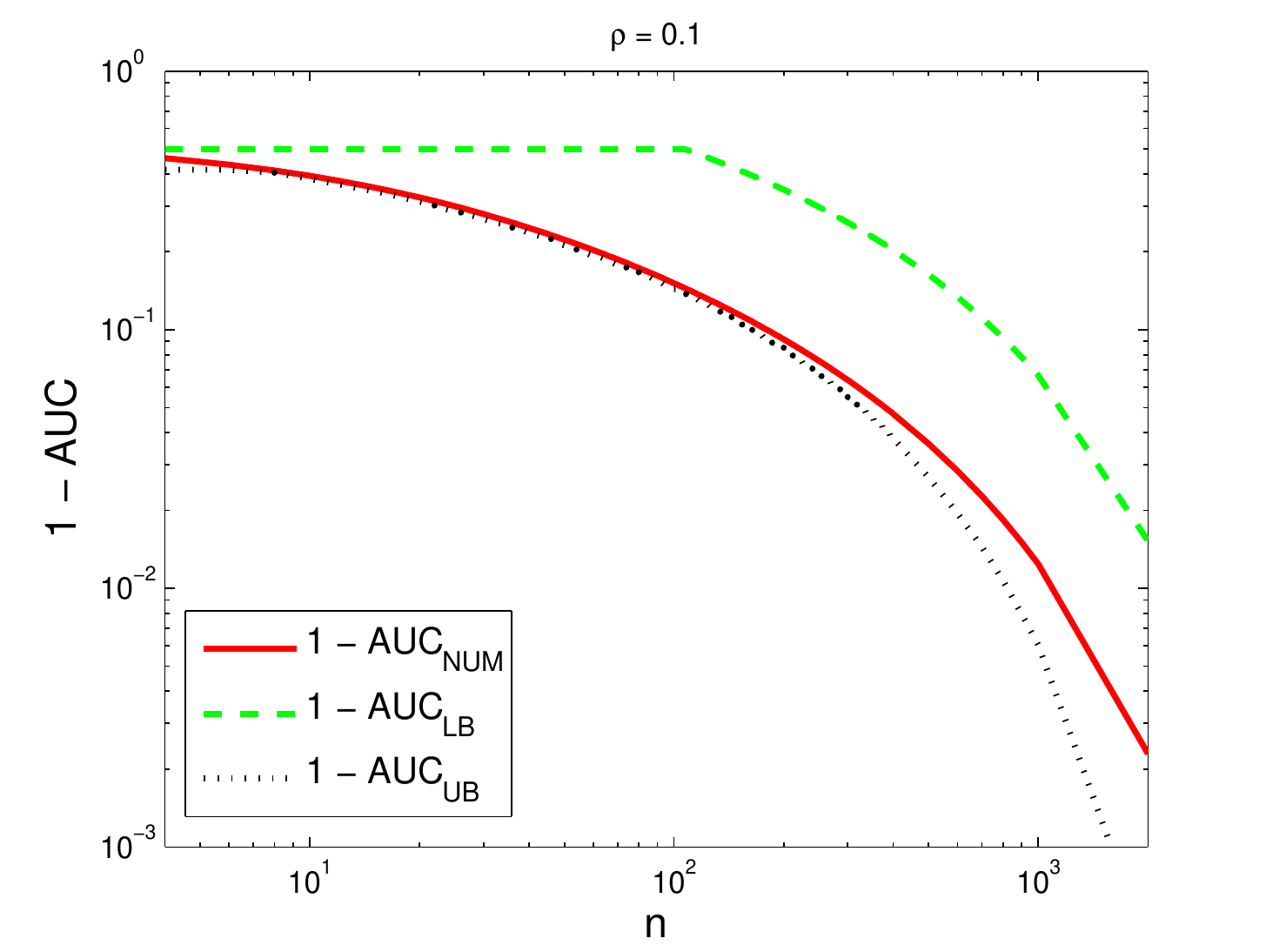}
\end{minipage}
\hspace{0.1cm}
\begin{minipage}[b]{0.48\linewidth}
\includegraphics[width=1.08\linewidth]{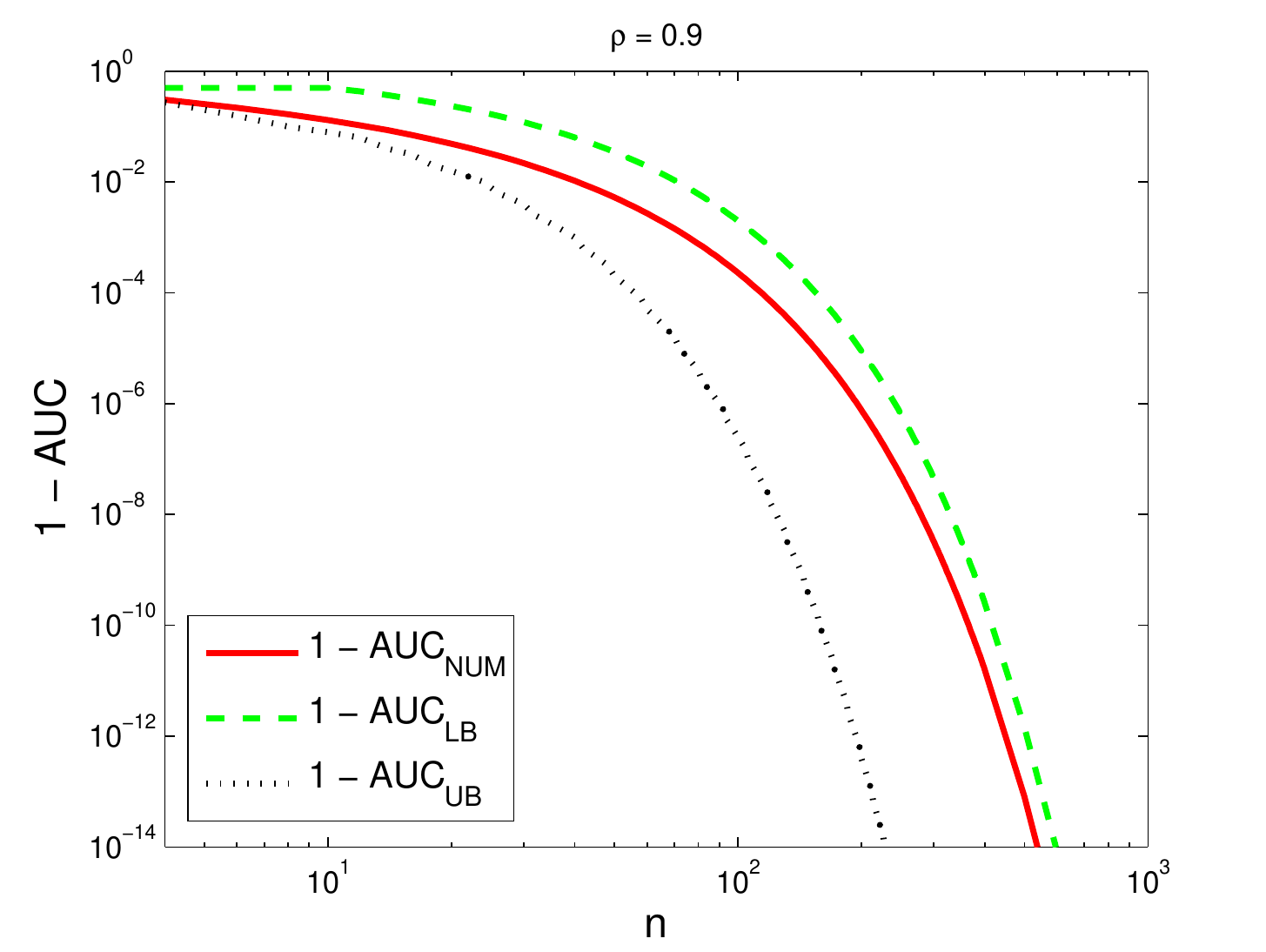}
\end{minipage}
\caption{$1-$AUC v.s. the dimension of the graph, $n$ for Star approximation of the Toeplitz example with $\rho = 0.1$ {\bf (left)} and $\rho=0.9$ {\bf (right)}. In both figures, the numerically evaluated AUC is compared with its bounds.}
\label{fig:Toeplitz_Example_Star}
\end{figure}

Figure \ref{fig:Toeplitz_Example_Star} plots the $1-$AUC v.s. the dimension of the graph, $n$ for different correlation coefficients, $\rho =0.1$ and $\rho=0.9$. This figure also indicates the upper bound and the lower bound for the $1-$AUC.

%
%
%
%

\subsubsection{Chain approximation}
The chain covariance matrix is as follows (nodes are connected like a first order Markov chain, $1$ to $n$)

\begin{equation*}
 \bSigma_{\uX_\mcT}^{chain} =
\begin{bmatrix}
                 1 & \rho & \rho^2 & \ldots  &   \rho^{n-1}     \\
             \rho & \ddots  & \ddots  & \ddots  &  \vdots  \\
             \rho^2 & \ddots & \ddots &  \ddots     & \rho^2  \\
             \vdots & \ddots & \ddots &  \ddots     & \rho  \\
             \rho^{n-1} &  \ldots & \rho^2 & \rho   &   1  \\
\end{bmatrix}
.
\end{equation*}
For this example, the KL divergence and the Jeffreys divergence can be computed in closed form as
$$\mcD (\uX||\uX_{chain}) = \mcD (\uX||\uX_{star}) $$
and
\begin{align*}
&\mcD_{\mcJ} (\uX \, , \, \uX_{chain})  =  \frac{\rho^2}{( 1 + (n-1) \rho )(1-\rho)} \times\\
&  \left( \frac{n(n-1)}{2}- \frac{n(1-\rho^n)}{1-\rho} + \frac{1-(n+1)\rho^n + n \rho^{n+1}}{(1-\rho)^2}  \right)
\end{align*}
respectively.
Moreover, for large values of $n$ we have the following approximation
$$\mcD_{\mcJ} (\uX \, , \, \uX_{chain}) \approx \frac{n}{2} \frac{\rho}{1 -\rho} .$$

\begin{figure}[ht]
\begin{minipage}[b]{0.48\linewidth}
\includegraphics[width=1.08\linewidth]{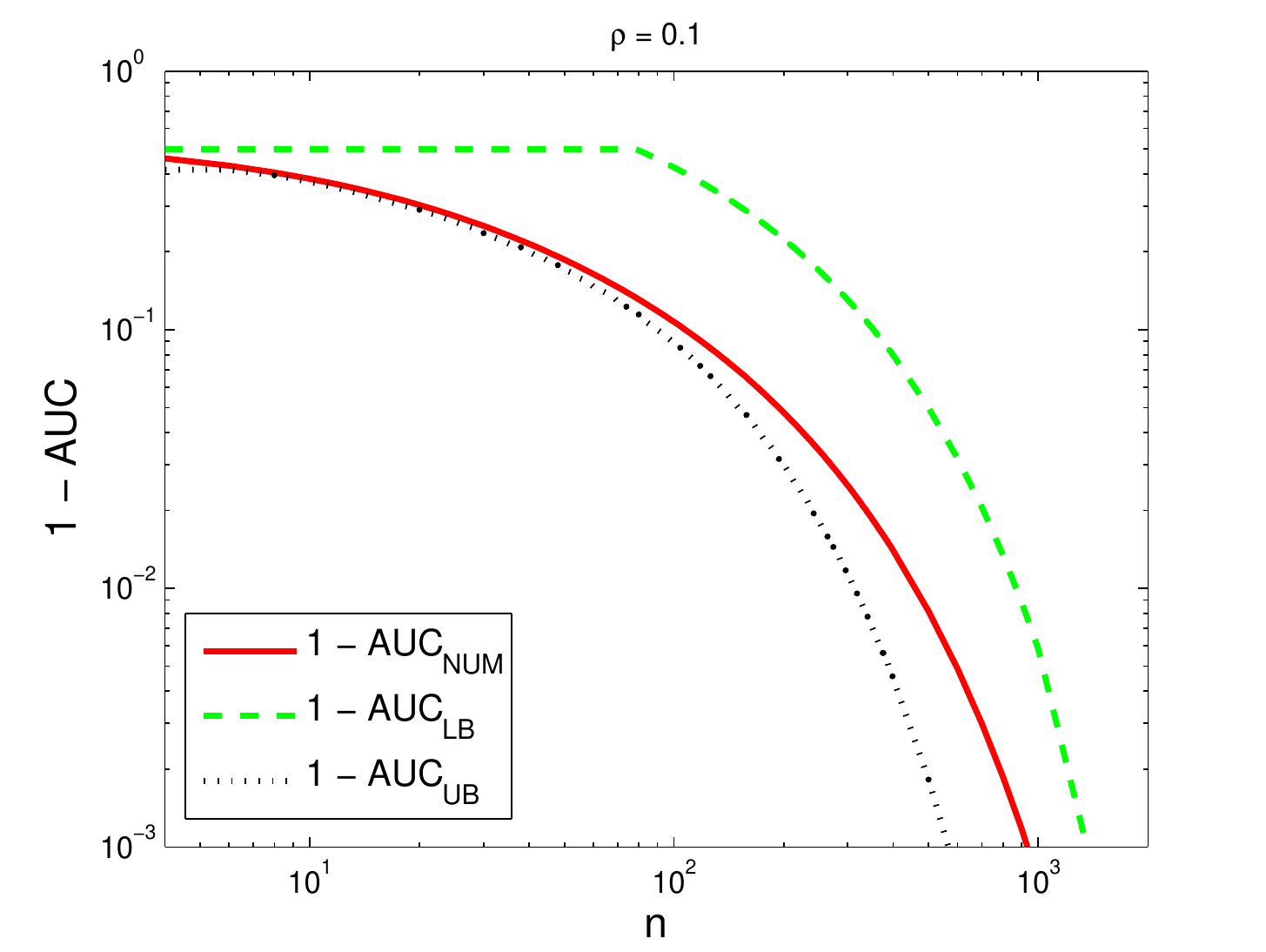}
\end{minipage}
\hspace{0.1cm}
\begin{minipage}[b]{0.48\linewidth}
\includegraphics[width=1.08\linewidth]{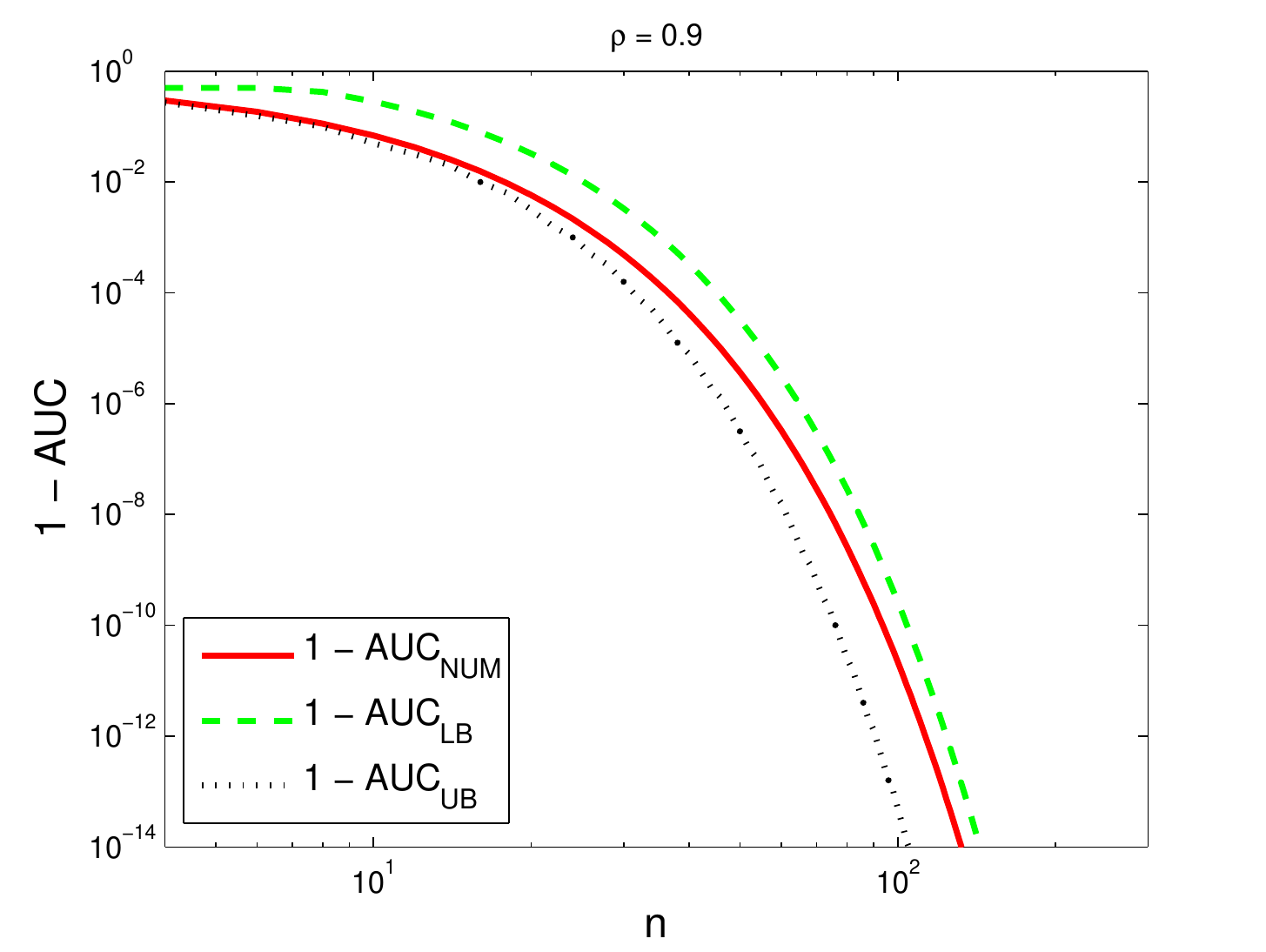}
\end{minipage}
\caption{$1-$AUC v.s. the dimension of the graph, $n$ for Chain approximation of the Toeplitz example with $\rho = 0.1$ {\bf (left)} and $\rho=0.9$ {\bf (right)}. In both figures, the numerically evaluated AUC is compared with its bounds.}
\label{fig:Toeplitz_Example_Chain}
\end{figure}

Figure \ref{fig:Toeplitz_Example_Chain} plots the $1-$AUC v.s. the dimension of the graph, $n$ for different correlation coefficients, $\rho =0.1$ and $\rho=0.9$ as well as its upper and lower bounds.

In both figure \ref{fig:Toeplitz_Example_Star} and figure \ref{fig:Toeplitz_Example_Chain}, $(1-\textnormal{AUC})$ and its bounds rapidly goes to $0$ which means that AUC goes to one as we increase the number of nodes, $n$, in the graph. 
More precisely, bounds for $1-\textnormal{AUC}$ are decaying exponentially as the dimension of the graph, $n$, increases which is consistent with the theory obtained for analytical bounds.
Furthermore, we can conclude from these figures that a smaller $\rho$ results in a better tree approximation, i.e. covariance matrices with smaller correlation coefficients are more like tree structure model.
Moreover, comparing the AUC for the star network approximation with the AUC for the chain network approximation we conclude that the star network is a much better approximation than the chain network even though that both approximation networks have the same KL divergences. 
We can also interpret this fact through the analytical bounds obtained in this paper. The star network is a better approximation than the chain network since the decay rate of $1-\textnormal{AUC}$ for the star network is less than its decay rate for the chain network.

\noindent {\bf Remark:} The star approximation in the above example has lower AUC than the chain approximation. Practically, it means the correlation between nodes that are not connected in the approximated graphical structure is more realistic in star network than the chain network.

\subsection{Solar data}

In this Example, covariance matrix is calculated based on datasets presented in \cite{APSIPA2014}.
Two datasets which are obtained from the National Renewable Energy Laboratory (NREL) website \cite{NREL}.
The first data set is the Oahu solar measurement grid which consists of $19$ sensors ($17$ horizontal sensors and two tilted sensors) and the second one is the NREL solar data for $6$ sites near Denver, Colorado. 
These two data sets are normalized using standard normalization method and the zenith angle normalization method \cite{APSIPA2014} and then the unbiased estimate of the correlation matrix is computed\footnote{See \cite{APSIPA2014} for fields definition and other details about the normalization methods for the solar irradiation covariance matrix.}.

\subsubsection{The Oahu solar measurement grid dataset}
\begin{figure}[ht]
\begin{minipage}[b]{0.48\linewidth}
\includegraphics[width=1.08\linewidth]{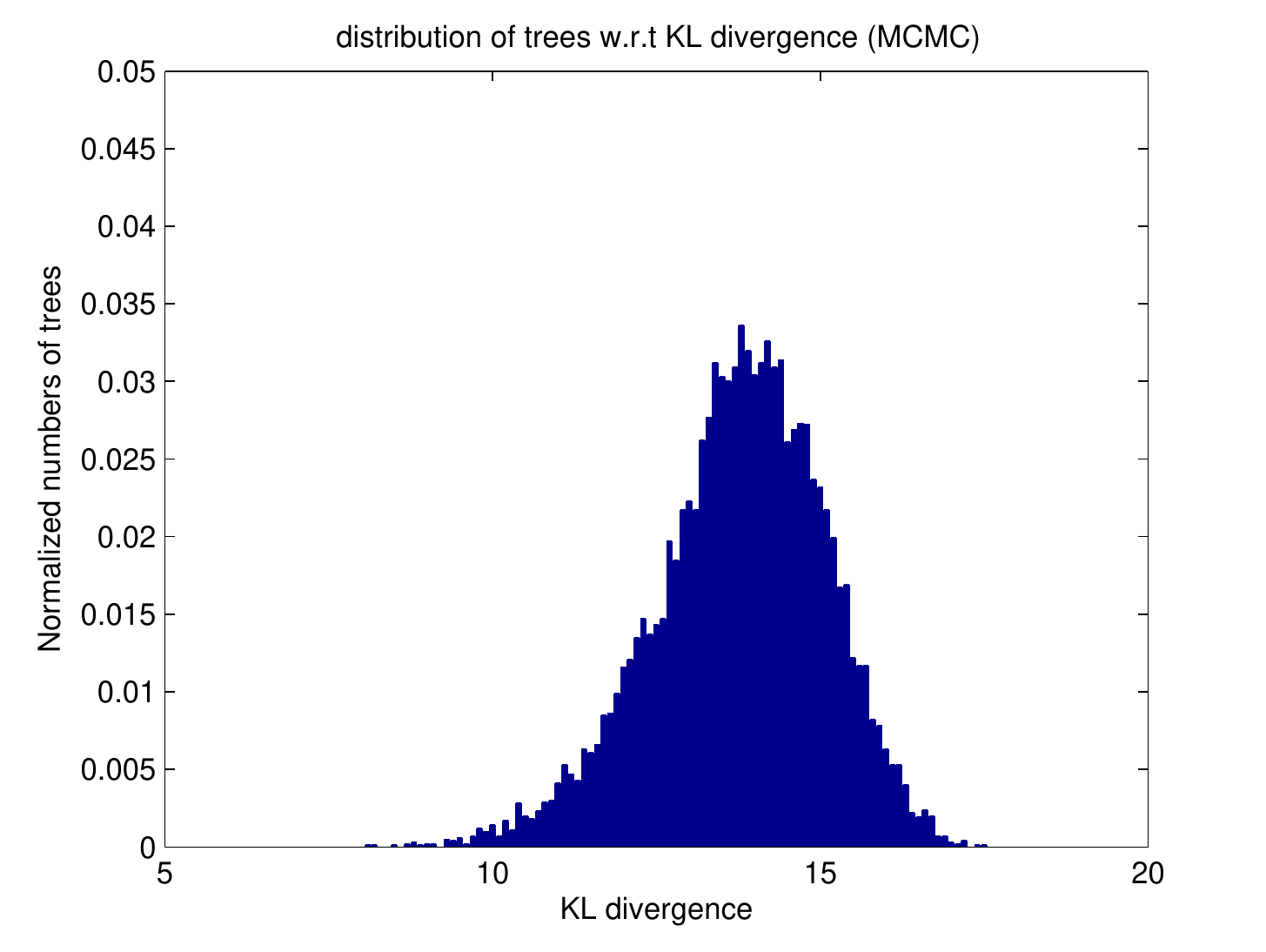}
\end{minipage}
\hspace{0.1cm}
\begin{minipage}[b]{0.48\linewidth}
\includegraphics[width=1.08\linewidth]{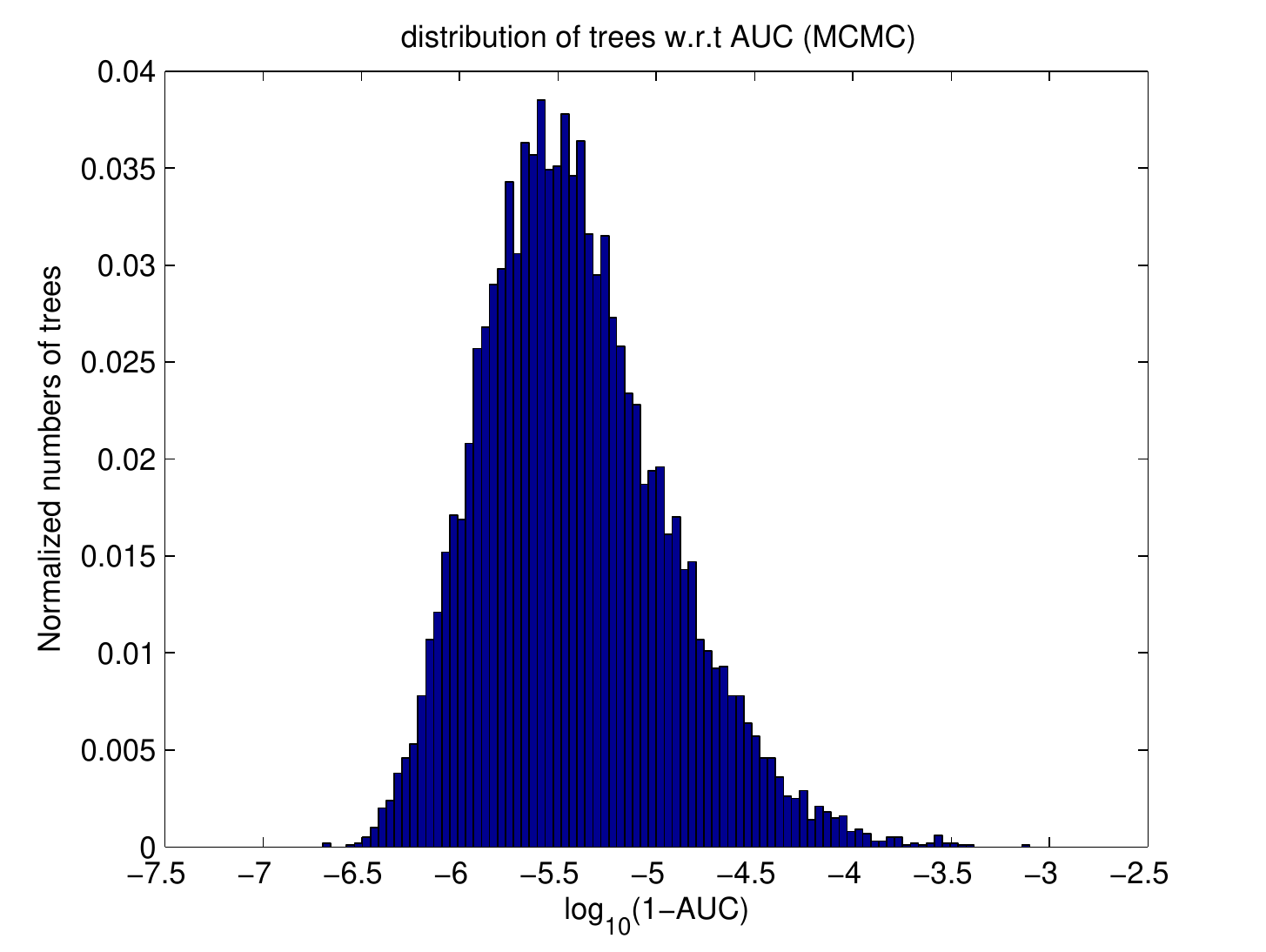}
\end{minipage}
\caption{{\bf Left:} distribution of the generated trees (Normalized histogram) using MCMC v.s. the KL divergence and {\bf Right:} distribution of the generated trees (Normalized histogram) using MCMC v.s. $\log_{10} (1-\textnormal{AUC})$ for the Oahu solar measurement grid dataset in summer season at 12:00 PM.}
\label{fig:Oahu_dataset}
\end{figure}

From data obtained from $19$ solar sensors at the island of Oahu, we computed the spatial covariance matrix during the summer season at 12:00 PM averaged over a window of 5 minutes. Then, the AUC and the KL divergence are computed for those tree structures that are generated using Markov Chain Monte-Carlo (MCMC) method. Figure \ref{fig:Oahu_dataset} shows the distribution of those tree structures generated using MCMC method versus the KL divergence {\bf (left)} and v.s $\log_{10} (1-\textnormal{AUC})$ {\bf (right)}\footnote{In this example, since the AUC for all generated tree structures is close to one, we plots the distribution of generated trees v.s. $\log_{10} (1-\textnormal{AUC})$.}.

Looking back at figure \ref{fig:FRLS}, for very small value of $1-\textnormal{AUC}$ the relationship between the KL divergence and the boundary of the possible feasible region for $-\log(1-\textnormal{AUC})$ is linear. This means that if the upper bound is tight then the relationship between the KL divergence and the $-\log(1-\textnormal{AUC})$ is almost linear.
In figure \ref{fig:Oahu_dataset}, the maximum value of $1-$AUC for this model is less than $10^{-3}$ which justifies why two distributions in figure \ref{fig:Oahu_dataset} are scaled/mirrored of each other.
Moreover, just by looking at the distribution of tree models in this example, it is obvious that most tree models have similar performance. Only a small portion of the tree models have better performance than the most trees, but the difference is not that significant.

\subsubsection{The Colorado dataset}
\begin{figure}[ht]
\begin{minipage}[b]{0.48\linewidth}
\includegraphics[width=1.08\linewidth]{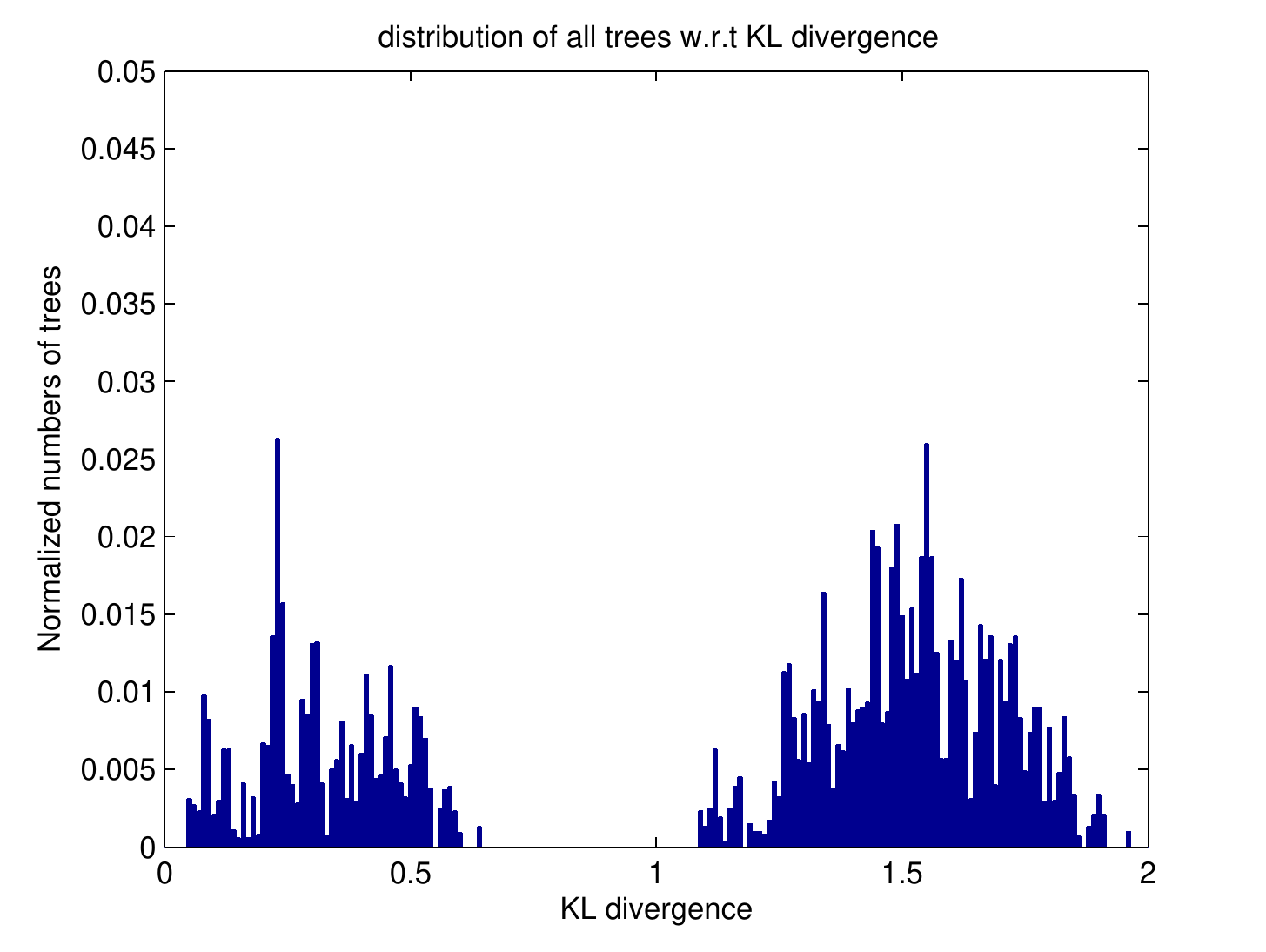}
\end{minipage}
\hspace{0.1cm}
\begin{minipage}[b]{0.48\linewidth}
\includegraphics[width=1.08\linewidth]{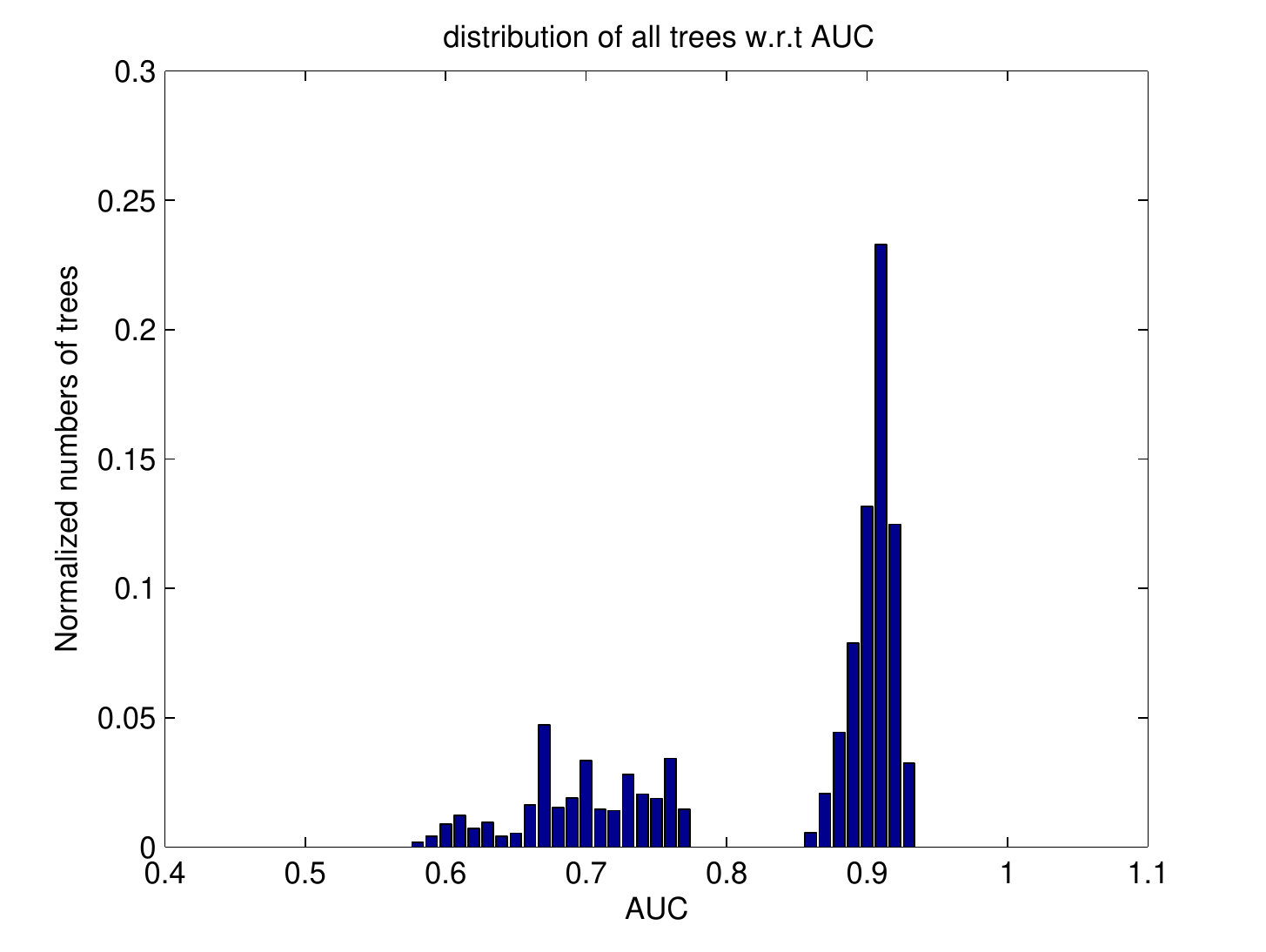}
\end{minipage}
\caption{{\bf Left:} distribution of all trees (Normalized histogram) v.s. the KL divergence and {\bf Right:} distribution of all trees (Normalized histogram) v.s. the AUC for the Colorado dataset in summer season at 12:00 PM.}
\label{fig:Colorado_dataset}
\end{figure}

From the solar data obtained from $6$ sensors near Denver, Colorado, we computed the spatial covariance matrix during the summer season at 12:00 PM averaged over a window of 5 minute. Then, the AUC and the KL divergence are computed for all possible tree structures. Figure \ref{fig:Colorado_dataset} shows the distribution of all possible tree structures v.s the KL divergence {\bf (left)} and v.s the AUC {\bf (right)}.

In the Colorado dataset, there are two sensors that are very close to each other compared to the distance between all other pairs of sensors.
As a result, if the particular edge between these two sensors is in the approximated tree structure we get a smaller AUC and KL divergence compared to when that particular edge is not in the tree structure. This explains why the distributions of all trees in this case looks like a mixture of two distributions.
This result also gives us valuable incite on how to answer the following question, "How to construct informative approximation algorithms for model selection in general." One catch as an example is that for the Colorado dataset, almost all trees that contain the particular edge between the two aforementioned sensors are good approximations while the rest of tree models' performances are not desirable.

\subsubsection{Two-dimensional sensor network}
\begin{figure}[ht]
\begin{minipage}[b]{0.48\linewidth}
\includegraphics[width=1.08\linewidth]{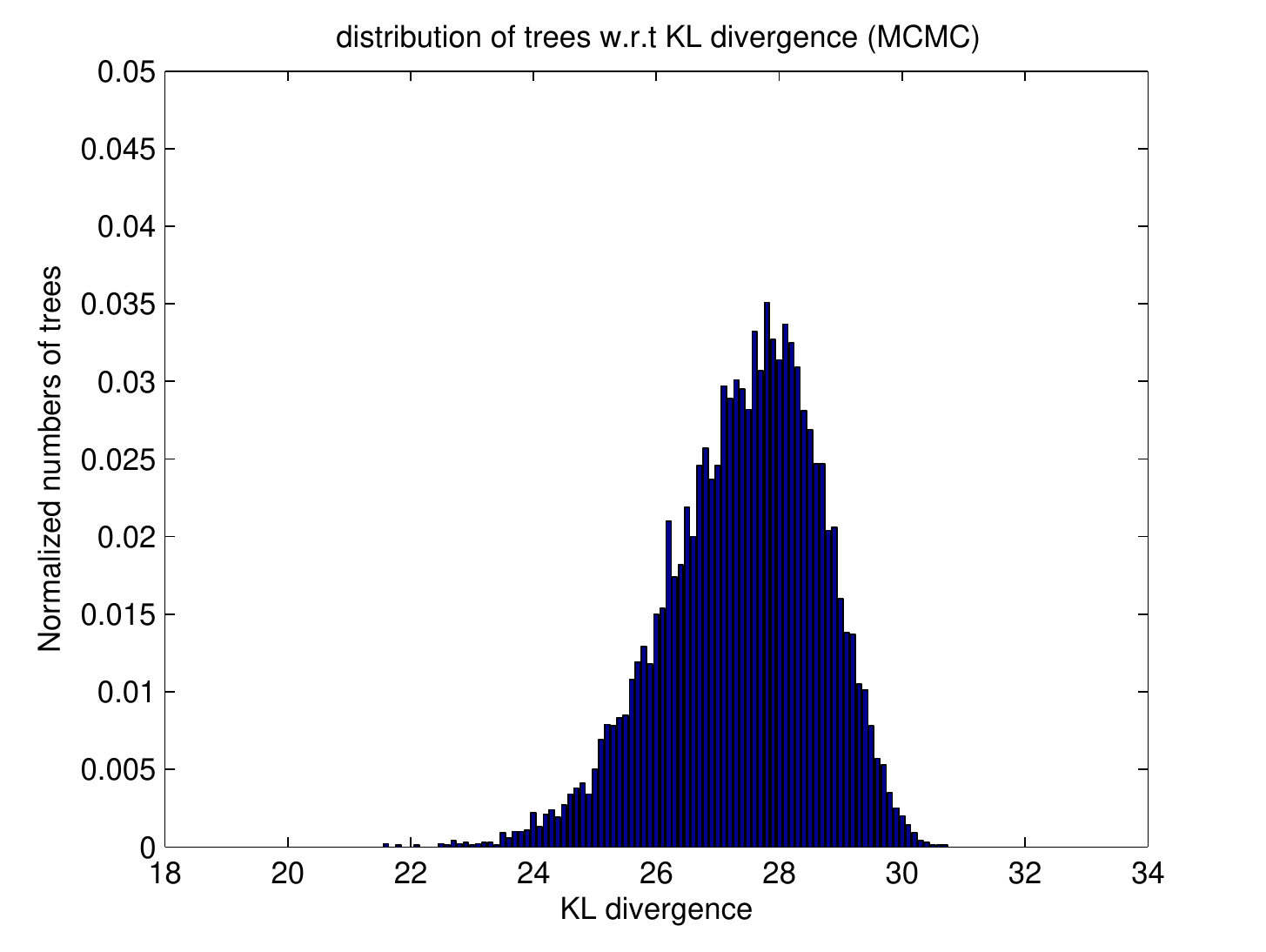}
\end{minipage}
\hspace{0.1cm}
\begin{minipage}[b]{0.48\linewidth}
\includegraphics[width=1.08\linewidth]{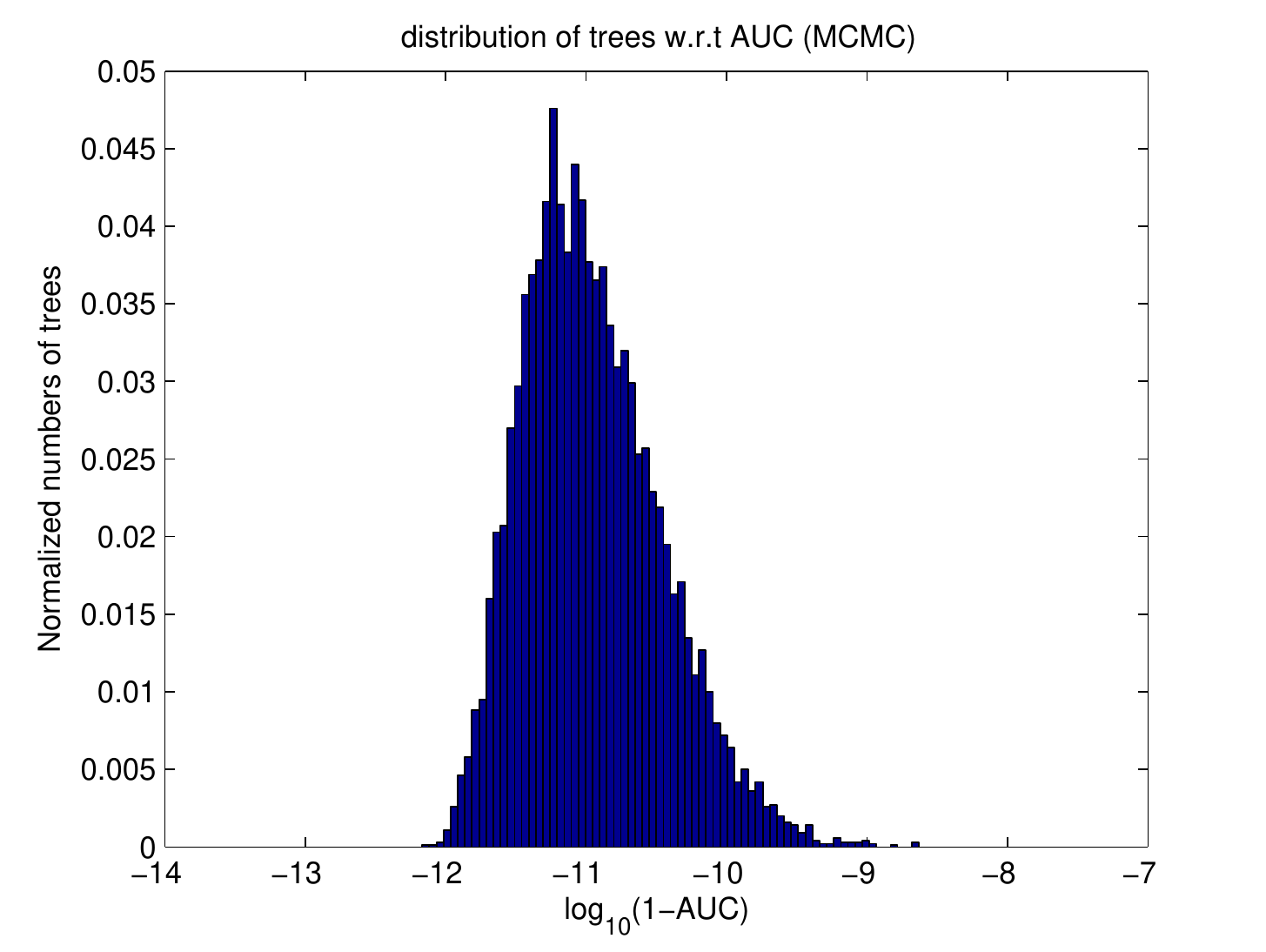}
\end{minipage}
\caption{{\bf Left:} distribution of the generated trees (Normalized histogram) using MCMC v.s. the KL divergence and {\bf Right:} distribution of the generated trees (Normalized histogram) using MCMC v.s. $\log_{10} (1-\textnormal{AUC})$ for the 2D sensor network example with $20$ sensors and $\sigma = 1$.}
\label{fig:2D_SN}
\end{figure}

\begin{figure}[ht]
\begin{minipage}[b]{0.48\linewidth}
\includegraphics[width=1.08\linewidth]{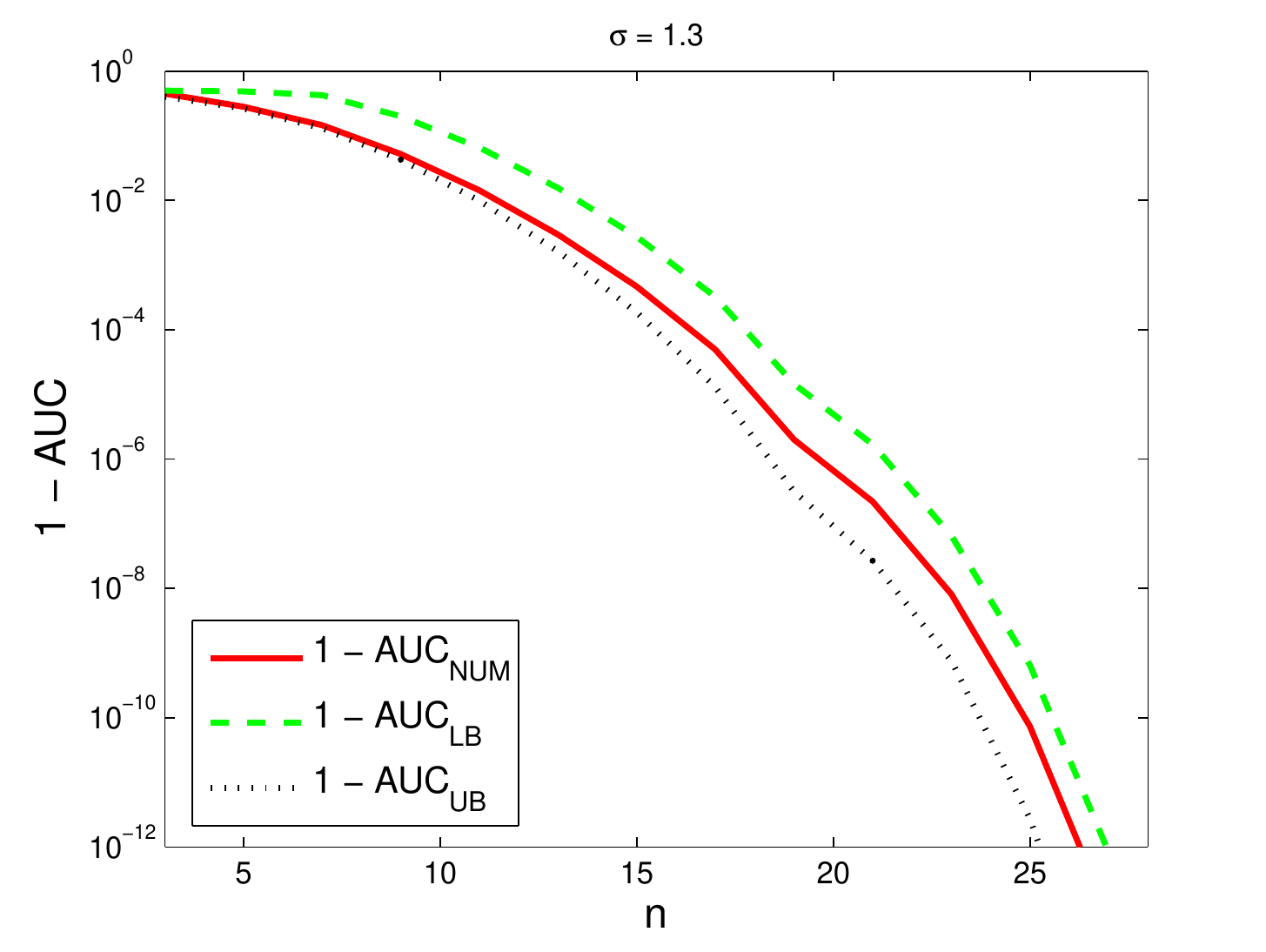}
\end{minipage}
\hspace{0.1cm}
\begin{minipage}[b]{0.48\linewidth}
\includegraphics[width=1.08\linewidth]{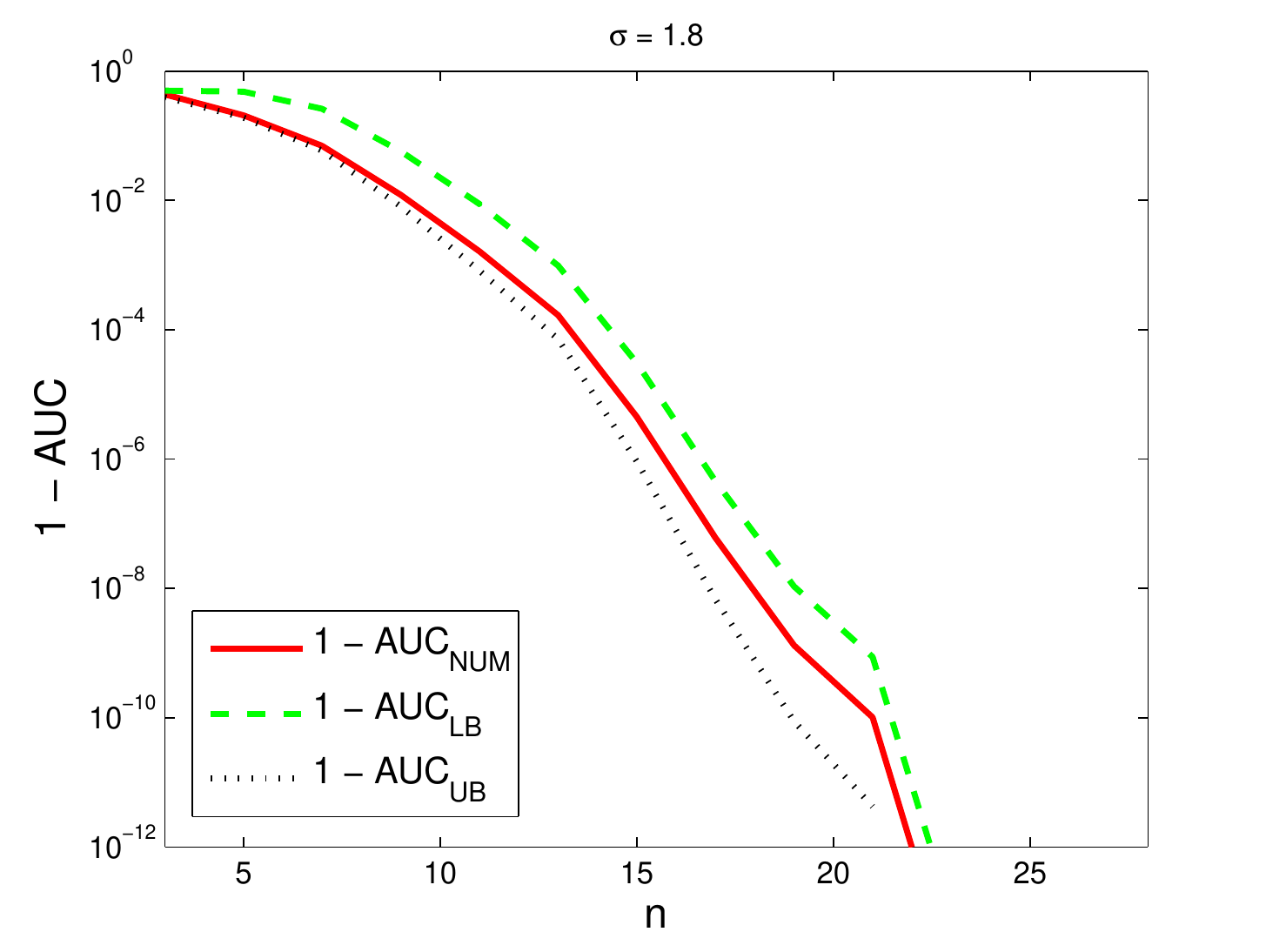}
\end{minipage}
\caption{  $1-$AUC and its bounds v.s. the dimension of the graph, $n$ for $\sigma = 1.3$ {\bf (left)} and $\sigma = 1.8$ {\bf (right)}, averaged over 1000 runs of sensor networks generated randomly.}
\label{fig:2D_SN_CB}
\end{figure}

In this example, we create a 2-dimensional (2D) sensor network using Gaussian kernel \cite{GaussianKernel} as follows
$$\Sigma_{\uX}(i,j) = \left[ e^{-\frac{d(i , j)^2}{2 \sigma^2}} \right]$$
where $d(i , j)$ is the Euclidean distance between the $i$-th sensor and the $j$-th sensor in the 2D space.
All sensors are located randomly in 2D space\footnote{Sensors location in each dimension are drawn randomly from a Normal distribution.}.
We set $\sigma = 1$ and generate a 2D sensor network with $20$ sensors.
For the 2D sensor network example, figure \ref{fig:2D_SN} shows the distribution of the generated tree structures using MCMC method v.s KL divergence {\bf (left)} and v.s $\log_{10} (1-\textnormal{AUC})$ {\bf (right)}.
Again we see the mirroring effect in Fig. \ref{fig:2D_SN} as we have an almost linear relationship between the KL divergence and $-\log (1-\textnormal{AUC})$.
Note that, the covariance matrix generated has one dominant eigenvalue in most cases.
Furthermore, figure \ref{fig:2D_SN_CB} plots $1-$AUC as well as its analytical upper bound and lower bound v.s. the dimension of the graph, $n$ for $\sigma = 1.3$ {\bf (left)} and $\sigma = 1.8$ {\bf (right)}. To generate this figure, we randomly generated $1000$ sensor networks and then plot the averaged AUC.
As we can see in this figure, the $1-$AUC and its bounds decay exponentially which is consistent with the theoretical results of this paper.

\section{conclusion}
\label{sec:con}

In this paper, we formulate a detection problem and investigate the quality of model selection problem. More specifically, we consider Gaussian distributions and discuss the covariance selection quality of a given model. We present the correlation approximation matrix (CAM), and show its relationship with information theory divergences such as the KL divergence, the reverse KL divergence and the Jeffreys divergence as well as the ROC curve and the area under it, i.e. the AUC, as a measure of accuracy in the detection problem framework. 
Moreover, this paper presents an analytical expression for the AUC that can efficiently be evaluated numerically. Also, the AUC analytical lower and upper bounds are provided in this paper. It is shown that the value of AUC and its bounds only depend on the eigenvalues of the CAM.
We pick the tree structure as an example of an approximation model and use the Chow-Liu MST algorithm to compute the maximum likelihood tree structure approximation.
Then, the quality of the Chow-Liu MST tree algorithm is investigated using the formulated detection problem.
Through some examples, we show that in general, the tree approximation is not a good model as the number of nudes in the graphical model increases which is the case in high dimensional problems such as {\it smart grid} and {\it big data}. 
The aforementioned result is also consistent with the analytical results provided in this paper that is $1-$AUC decays exponentially as the dimension of graph increases.

The detection framework presented in this paper, can be generalized for non-Gaussian models.
Moreover, the AUC analytical bounds obtained in this paper can also be used in other applications that are using AUC as a relevant criterion. One example is in medicine when the AUC is used for diagnostic tests between positive instance and negative instance \cite{AUCinMed} where instead of changing the coordinates we can look at the exponent of the AUC bounds.
\textcolor{black}{In future work, we will define symmetric CAM and show that model selection problems such as tree approximation can be see as a linear transformation}

\section*{Acknowledgment}

\textcolor{black}{Authors would like to thank Prof. Peter Harremo{\"e}s for his helpful discussions on information divergences and assistance with Theorem \ref{thm:ART}. }
This work was supported in part by NSF grant ECCS-1310634, the Center for Science of Information (CSoI), an NSF Science and Technology Center, under grant agreement CCF-0939370, and the University of Hawaii REIS project.

\bibliographystyle{IEEEbib}
\bibliography{refs}

\appendices

\section{Proof of Lemma \ref{lem:RKL}}
\label{apx:RKL}
The calculus based proof for the special case of continuous PDFs is as follow. We can apply the Leibniz integral rule \cite{LeibnizRule} and compute the derivative of CDFs $P_0(l)$ and $P_1(l)$ as
$$f_{L_0}(l) = - \frac{d P_0(l)}{d l}$$
and
$$f_{L_1}(l) = - \frac{d P_1(l)}{d l}$$
since $f_{L_0}(l)$ and $f_{L_1}(l)$ are continuous functions.\footnote{Both $f_{L_0}(l)$ and $f_{L_1}(l)$ are PDFs in generalized Chi-squared distributions class. This means that each of these PDFs are convolution of weighted Chi-squared distributions. Weighted Chi-squared distribution is continuous in its domain thus, convolution of these distributions is continuous in its domain.}
We have
\begin{align*}
\mcD \left(f_{L_0}(l) || f_{L_1}(l)\right) & =  \int_{-\infty}^{+\infty} \log \frac{f_{L_0}(l)}{f_{L_1}(l)} \; f_{L_0}(l) \, dl \\
& \stackrel{(a)}{=}    - \int_{0}^{1}  \log \frac{d P_1}{d P_0} \, d P_0 \\
& \stackrel{(b)}{=} - \int_{0}^{1} \log h'(z) \, dz
\end{align*}
where equality (a) is true since we can replace PDfs $f_{L_0}(l)$ and $f_{L_1}(l)$ using the derivative of their CDFs.
Equality (b) is just a change of variable, $z =  P_0(l)$, in order to write the integral in terms of the derivative of the ROC curve. 
Proof for the second part of this lemma is similar to the proof of the first part.
\hfill \ensuremath{\blacksquare}

\section{Proof of Theorem \ref{thm:ART}}
\label{apx:LMS}

Looking back at properties of the ROC curve, $h(z)$, where $z \in [0,1]$, the ROC curve have to satisfy the following conditions
\begin{itemize}
\item {\bf C1:} $\int_{0}^{1} h'(z) dz = 1$
\item {\bf C2:} $h'(z) \geq 0$
\item {\bf C3:} $h'(z)$ is decreasing
\end{itemize}
where $h'(z)$ is the derivative of the ROC curve, $h(z)$.
Also for a given ROC curve, $h(z)$, we can compute the AUC as 
\begin{equation*}
\tPr \left( L_{\Delta}>0 \right) = \int_{0}^{1}  h(z) dz.
\end{equation*}
Then, using integration by parts, we can show that
\begin{equation*}
1 - \tPr \left( L_{\Delta}>0 \right) = \int_{0}^{1} z \, h'(z) dz.
\end{equation*}

To compute the possible feasible region stated in the theorem \ref{thm:ART}, we need to optimize both of following KL divergences, $\mcD \left(f_{L_1}(l) || f_{L_0}(l)\right)$ and $\mcD \left(f_{L_0}(l) || f_{L_1}(l)\right)$, with respect to the derivative of the ROC curve given a fixed AUC, $\tPr \left( L_{\Delta}>0 \right)$, while conditions, C1, C2 and C3 hold. To solve this optimization, we can use the method of Lagrange multiplier. 

{\bf First step:} Here we minimize $\mcD \left(f_{L_1}(l) || f_{L_0}(l)\right)$ with respect to the derivative of the ROC curve given the constraints. Optimization problem is as follow
\begin{align}
\label{eq:optKL10}
\argmin_{h'(z)} \quad & \quad - \int_{0}^{1}  \log h'(z) dz \\\notag
\textnormal{s. t. } & \quad \int_{0}^{1} z \, h'(z) dz = 1 - \tPr \left( L_{\Delta}>0 \right) \\\notag
& \quad \textnormal{C1, C2 \& C3. }
\end{align}
To solve this optimization problem, we first write the Lagrangian. 
We need two coefficients $a$ and $b$ corresponding to conditions in optimization problem \eqref{eq:optKL10}. Then, we can write the Lagrange multiplier as a function of the derivative of the ROC curve, $z$, $a$ and $b$ as follow
\begin{align*}
L( h'(z) , z , a , b ) & = - \int_{0}^{1} \log h'(z) dz \\
& + a \left( \int_{0}^{1} z h'(z) dz - \left( 1- \textnormal{Pr} \left( L_{\Delta}>0 \right) \right) \right) \\
& + b \left( \int_{0}^{1} h'(z) dz - 1 \right).
\end{align*}
Note that, the Lagrangian, $L( h'(z) , z , a , b ) $ is a convex function of $h'(z)$.
Thus, we can compute its minimum by taking its derivative with respect to $h'(z)$. Doing so, we get
$$\frac{\partial L( h'(z)  , z , a , b )}{\partial h'(z)} = \int_{0}^{1} \left( az+b - \frac{1}{h'(z)} \right) dz.$$
Set $\frac{\partial L( h'(z) , z , a , b ) }{\partial h'(z)} = 0$ we get
$$ h'(z) = \frac{1}{az+b}$$
for all $z \in [0,1].$
From C3, since $h'(z)$ is decreasing, we can conclude that $a > 0$.
Moreover, from C1, at optimum we have $\int_{0}^{1} h'(z) dz = 1$ and thus, we can compute one of the coefficients as $b = \frac{a}{e^{a} -1}$.

Computing the AUC integral and the KL divergence using the ROC curve we get the following parametric boundary for the possible feasible region
\begin{equation}
\label{eq:B1}
\tPr \left( L_{\Delta}>0 \right) = \frac{1}{1-e^{-a}} - \frac{1}{a}
\end{equation}
and
\begin{equation}
\label{eq:B2}
\mcD = \log(a) +  \frac{a}{e^{a} - 1} - 1 - \log(1-e^{-a})
\end{equation}
where $\mcD = \mcD \left(f_{L_1}(l) || f_{L_0}(l)\right)$.

{\bf Second step:} Here we minimize $\mcD \left(f_{L_0}(l) || f_{L_1}(l)\right)$. The Lagrange multiplier for this step is similar to the first step but it is more straight forward if we define $g(\eta)=h^{-1}(\eta)$. Note that using integration by parts, we can show that AUC is 
\begin{equation*}
\tPr \left( L_{\Delta}>0 \right) = \int_{0}^{1} \eta \, g'(\eta) d\eta.
\end{equation*}
Now, we can write the Lagrangian for the optimization problem with respect to $g'(\eta)$. The Lagrangian is convex with respect to $g'(\eta)$, thus taking the derivative and set it equal to zero as follow
$$\frac{\partial L( g'(\eta) , \eta , a , b )}{\partial g'(\eta)} = 0$$
we can compute the parametric boundary for the possible feasible region.
The parametric boundary in this case is the same as solution in \eqref{eq:B1} and \eqref{eq:B2} with $\mcD = \mcD \left(f_{L_0}(l) || f_{L_1}(l)\right)$. 
Thus, combining these two steps, for the optimal boundary we have
$$\mcD^{*}_l = \min \{ \mcD \left(f_{L_1}(l) || f_{L_0}(l)\right) \, , \, \mcD \left(f_{L_0}(l) || f_{L_1}(l)\right) \}.$$
\hfill \ensuremath{\blacksquare}

\end{document}